\documentclass[
 reprint,
 superscriptaddress,
 altaffilletter,
 amsmath,amssymb,
 aps,
 prd,
 longbibliography
floatfix,
]{revtex4-1}

\pdfoutput=1 

\usepackage{graphicx}
\usepackage{float}
\usepackage[T1]{fontenc}
\usepackage[version=4]{mhchem}
\usepackage{wasysym}
\usepackage{siunitx}
\usepackage{xspace}
\usepackage{multirow}
\usepackage[normalem]{ulem}
\usepackage[colorlinks=true]{hyperref}
\usepackage[usenames, dvipsnames, table]{xcolor} 
\usepackage{placeins}
\usepackage[ignoreunlbld,nocites,norefs]{refcheck}

\definecolor{mygreen}{rgb}{0.0, 0.5, 0.0}
\definecolor{myblue}{rgb}{0.0, 0.0, 0.5}

\newcommand{\0}{\ensuremath{0\nu\beta\beta}}
\newcommand{\2}{\ensuremath{2\nu\beta\beta}}

\newcommand{\Q}{\ensuremath{Q_{\beta\beta}}}

\usepackage[draft, inline, nomargin, final]{fixme}
\fxsetup{theme=color,mode=multiuser}
\FXRegisterAuthor{mh}{anmh}{\color{red}MH}	
\FXRegisterAuthor{jb}{anjb}{\color{orange}JB}
\FXRegisterAuthor{lk}{anlk}{\color{myblue}LK} 
\FXRegisterAuthor{ss}{anss}{\color{blue}SS} 
\FXRegisterAuthor{ts}{ants}{\color{green}ts}

\begin{document}

\title{Sensitivity and discovery potential of the proposed nEXO experiment \\ to neutrinoless double-$\beta$ decay}

\author{J.~B.~Albert}
\affiliation{Department of Physics and CEEM, Indiana University, Bloomington, Indiana 47405, USA}
\author{G.~Anton}
\affiliation{Erlangen Centre for Astroparticle Physics (ECAP), Friedrich-Alexander University Erlangen-N\"urnberg, Erlangen 91058, Germany}
\author{I.~J.~Arnquist}
\affiliation{Pacific Northwest National Laboratory, Richland, Washington 99352, USA}
\author{I.~Badhrees}
\affiliation{Department of Physics, Carleton University, Ottawa, Ontario K1S 5B6, Canada}
\author{P.~Barbeau}
\affiliation{Department of Physics, Duke University, and Triangle Universities Nuclear Laboratory (TUNL), Durham, North Carolina 27708, USA}
\author{D.~Beck}
\affiliation{Physics Department, University of Illinois, Urbana-Champaign, Illinois 61801, USA}
\author{V.~Belov}
\affiliation{Institute for Theoretical and Experimental Physics, Moscow, Russia}
\author{F.~Bourque}
\affiliation{Universit\'e de Sherbrooke, Sherbrooke, Qu\'ebec J1K 2R1, Canada}
\author{J.~P.~Brodsky}
\affiliation{Lawrence Livermore National Laboratory, Livermore, California 94550, USA}
\author{E.~Brown}
\affiliation{Department of Physics, Applied Physics and Astronomy, Rensselaer Polytechnic Institute, Troy, New York 12180, USA}
% ...

\collaboration{nEXO Collaboration}
\noaffiliation

\date{May 1, 2018}

\begin{abstract}
The next-generation Enriched Xenon Observatory (nEXO) is a proposed experiment to search for neutrinoless double beta (\0) decay  in \ce{^136Xe} with a target half-life sensitivity of approximately $10^{28}$ years using $5\times10^3$ kg of isotopically enriched liquid-xenon in a time projection chamber.  This improvement of two orders of magnitude in sensitivity over current limits is obtained by a significant increase of the \ce{^136Xe} mass, the monolithic and homogeneous configuration of the active medium, and the multi-parameter measurements of the interactions enabled by the time projection chamber. The detector concept and anticipated performance are presented based upon demonstrated realizable background rates.
\end{abstract}

\maketitle
\flushbottom

\section{Introduction}

The observation of neutrinoless double-$\beta$ (\0) decay  would reveal fascinating new physics by demonstrating lepton number violation and confirming the existence of elementary Majorana fermions \cite{Majorana:1937}. This could impact our understanding of the neutrino mass generation mechanisms and may help illuminate possible origins of the cosmic baryon asymmetry \cite{Fukugita:1986hr}. Potential for discovery underpins the motivation for pursuing the ambitious experimental program described in this article.

Indeed, this is the same type of discovery quest that drives the construction of large accelerators to probe higher energies or that triggers the construction of larger telescopes capable of peering deeper in to the universe.
This exciting opportunity has gained widespread interest following the discovery of neutrino oscillations \cite{Fukuda:1998mi,Ahmad:2002jz,Eguchi:2002dm} that prove neutrino masses are not all zero.  Indeed, the difference between Majorana and Dirac neutrinos is only observable in the case of nonzero neutrino mass.  

While discovery of Majorana neutrinos and lepton-number violation is the principal goal, the discovery of \0 decay provides useful information on neutrino mass values, even if systematic uncertainties deriving from nuclear physics and the particular mechanism responsible for the decay obscure the translation from a measured half-life into an effective Majorana mass.
In this context, \0 discovery requirements are often represented in the parameter space of the  effective Majorana neutrino mass $\langle m_{\beta\beta}\rangle$ versus the mass of the lightest neutrino eigenstate, where 
\begin{equation}
\langle m_{\beta\beta}\rangle = \left| \sum_{i=1}^{3} U^2_{ei}m_i \right|
\end{equation}
and $m_i$ are the neutrino mass eigenvalues and $U_{ei}$ the mixing matrix elements.
The experimentally observable \0 decay half-life is inversely proportional to $\langle m_{\beta\beta}\rangle^2$:
\begin{equation}
[T^{0\nu}_{1/2}]^{-1} = \frac{\langle m_{\beta\beta}\rangle^2}{m^2_e}G^{0\nu}|M^{0\nu}|^2 ,
\label{eq:T-m}
\end{equation}
where $m_e$ stands for the electron mass, $G^{0\nu}$ is the phase space factor, and $M^{0\nu}$ the nuclear matrix elements.
This treatment assumes that \0 is mediated by the exchange of a light Majorana neutrino \cite{Avignone:2007fu}. While the effective Majorana mass is useful for the comparison of sensitivities of experiments studying different nuclides, care must be taken to account for uncertainties in the axial-vector coupling constant and nuclear matrix elements.
Recent work \cite{Agostini:2017jim,Caldwell:2017mqu} tried to quantify the \0 discovery potential for different nuclides and starting assumptions.

Among the candidate double-$\beta$-decaying nuclei \cite{PDG2016}, \ce{^136Xe} is particularly attractive. The \0 $Q$ value is high at \Q$=2458.07\pm0.31$ keV \cite{PhysRevLett.98.053003, McCowan:2010zz}. Xenon reserves in the atmosphere are practically unlimited and commercial production is sufficient to support large scale experiments. As a noble gas, isotopic enrichment can be performed efficiently. 
Xenon is readily purified of chemical  contaminants, resulting in exceedingly low intrinsic radioactive backgrounds \cite{Albert:2015nta,Albert:2014awa} and very long electron life-time \cite{exo-200_detector_paper,Albert:2013gpz}. As a radiation detector medium, liquid xenon shows high scintillation and charge yield \cite{Aprile:2009dv}. The high density ($\sim3$ g/cm$^3$) and high atomic number ($Z=54$) result in short attenuation lengths for $\gamma$ radiation. Finally, the xenon supply can be recycled into upgraded detectors as technology improves.  

The current most stringent limit on \0 decay in \ce{^{136}Xe} is $T_{1/2}^{0\nu} > 1.07 \times 10^{26}$ yr at 90\% C.L. from the KamLAND-Zen experiment \cite{KamLAND-Zen:2016pfg}, with a corresponding sensitivity of $5.6\times10^{25}$ yr from 504~kg$\cdot$yr exposure. The diagnostic power of the liquid xenon (LXe) time projection chamber (TPC) allows the EXO-200 experiment \cite{exo-200_detector_paper}, with a substantially smaller exposure of 178 kg$\cdot$yr, to reach a competitive sensitivity of $3.7 \times 10^{25}$ yr \cite{Albert:2017owj}.

nEXO, a follow-on to the EXO-200 program, is a proposed next-generation \0 experiment utilizing 5000 kg of isotopically enriched liquid xenon in a TPC. 
The TPC's multiparameter measurement capability allows the simultaneous determination of event energy, position, site multiplicity, and particle type. This capability, combined with the use of a large homogeneous detector volume, allows the optimal determination of the \0 signal and backgrounds while utilizing the entire xenon mass. An inherent low-background detector design results from the deliberative evaluation and choice of materials, underground location, and the use of a layered scheme of passive and active shielding. 

The multiparameter measurement of each event is an essential feature of a homogeneous detector with dimensions (of order of 100 cm) substantially larger than the attenuation length (of order of 10 cm) of $\gamma$-rays of energy similar to the \0 decay $Q$-value. 
The multiparameter event signature, allowing us to ``resolve'' a signal in more than one parameter, and the option of validating an observation through running with natural xenon, makes nEXO unique in its ability to discover a \0 signature.
Measurement by means of multivariable observations going beyond a one-dimensional peak search is an essential aspect of any believable discovery. This important point is often overlooked when experiments are discussed in terms of a one-dimensional rate analysis, where a low background is the only tool to perform a sensitive search. While low background is still very important, a logical question to ask is: how low does it need to be if several attributes can be determined simultaneously?
In particular, the power of a multiparameter experiment such as nEXO may substantially reduce risk, by allowing for a design that involves material purities already demonstrated at the time of the detector design.

In this article we present the experiment design, and predict the sensitivity reach and discovery potential of nEXO, explicitly taking into account its multiparameter event structure.

A brief review of the EXO-200 detector is provided in Sec.~\ref{sec:EXO-200}. This is important because the success of EXO-200 does more than validate the generic concept of nEXO. In fact, the background in nEXO and the sensitivity discussed here are, for the most part, projected using materials whose radioactive contaminations were already tested in EXO-200. The larger detector size and more advanced instrumentation are the ingredients required to obtain the sensitivity presented here.

Section~\ref{sec:nexo_design} introduces the conceptual design of the experiment, highlighting the novel solutions that are being investigated to enhance performance and deal with the size of a tonne-scale experiment.

Establishing a background model for the sensitivity calculation requires a careful assessment of the various contributions, in particular those from radioactivity in the detector materials. nEXO's approach to building the background model is based on the successful methodology developed by EXO-200 and is grounded in existing radio-assay measurements and Monte Carlo particle transport. A complete and quantitative evaluation of potential sources of background is described in Sec.~\ref{sec:backgrounds}.
The simulations that are used to model the detector and background are introduced in Sec.~\ref{sec:simulation}.

A discussion of the sensitivity methodology, which is based on the frequentist approach, is presented in Sec.~\ref{sec:sensitivity}. 
This is followed in Sec.~\ref{sec:results} by the expected background budget and results for sensitivity and discovery potential.

\section{EXO-200 Experiment}
\label{sec:EXO-200}

EXO-200 is an ongoing, 100 kg-class double-$\beta$-decay experiment, conceived around 2005.  It was envisaged as a tool for $\beta\beta$ discovery, but also as a test bench to develop the technology of a large, ultra-low background tracking calorimeter based on a TPC,  utilizing isotopically enriched Xe. In many respects EXO-200 serves as a successful prototype for nEXO.
A detailed technical description of the EXO-200 detector has been published~\cite{exo-200_detector_paper}.

EXO-200 has a total LXe mass of 175 kg held at a temperature of 167 K and isotopically enriched to a $^{136}$Xe abundance of 80.7\% \cite{Albert:2013gpz}. 
Thanks to a LXe density of 3.03 g/cm$^3$,
EXO-200 is compact, reducing the amount of low-activity materials needed to construct the device, when compared to a gas-phase detector. 

A TPC, designed for simultaneous charge and light read-out, forms the heart of the experiment. The EXO-200 collaboration was the 
first to employ combined light and charge read-out in a LXe TPC exploiting the anti-correlation of these two energy deposition processes \cite{exo-200_anti_correlation}. 
EXO-200 combined read-out of these two energy deposition processes allows full three-dimensional spatial reconstruction of energy deposits, and results in an energy resolution of $\rm \sigma/Q_{\beta\beta}=1.23\; \%$ \cite{Albert:2017owj}.

EXO-200 is a double-sided cylindrical TPC, with the high voltage cathode
in the middle and charge and light read-out planes on either end, perpendicular to the cylinder's axis. Each charge read-out plane consists of two sets of wires oriented at 60 degrees with respect to each other and a read-out pitch of 9 mm. 
The 175 nm Xe scintillation light is measured by an array of Avalanche Photo Diodes (APDs) on either end. The TPC body is made from thin electrolytic copper specially produced  by Aurubis of Hamburg (Germany) to minimize the concentrations levels of naturally occurring radioactive impurities. Copper handling and storage limited its exposure to cosmic radiation.

The TPC is submerged in a bath of 4140 kg of HFE-7000~\cite{HFE-7000}, an engineered 3M cryogen that remains liquid at the operating temperature. The HFE-7000 further serves as the innermost ultraradiopure radiation shield. The HFE is contained in a double-walled, vacuum insulated copper cryostat. 
The xenon is continuously extracted from the TPC, evaporated, circulated through a purifier, and recondensed to achieve an electron lifetime greater than 2 ms. The detector is installed underground at the Waste Isolation Pilot Plant (WIPP), near Carlsbad NM (USA).

The experience gained during the development and construction of EXO-200 forms one of the cornerstones of the nEXO experiment design. Background estimation and reduction play a particularly prominent role in this process.
The large EXO-200 data set allows for a detailed understanding of the composition of the background that has actually been achieved. 
The comparison of a data-driven EXO-200 background model to the pre-data taking 
projections can be interpreted as an estimate of the systematic uncertainties
inherent to the approach chosen during EXO-200 construction and envisaged for nEXO.

The pre-data taking assessment of the EXO-200 background is documented in reference~\cite{exo-200_detector_paper}. This assessment was based on the
determination of the radioactivity content of all detector components, coupled with a \textsc{Geant3.21} based detector simulation to compute event rates. 
All detector components went through a detailed and comprehensive material screening and selection program~\cite{exo-200_background_2008, Leonard:2017okt}.
The EXO-200 collaboration adopted a background goal 
for events near the \0 $Q$ value, 
which in turn determined the allowable $^{60}$Co, $^{232}$Th and $^{238}$U content of the detector. 
A similar approach was used to ensure that other background sources like external $\gamma$s, outgassed \ce{^222Rn}, muons, and cosmogenic \ce{^60Co} and \ce{^137Xe} remained within allowable limits. Shield design, and material screening and handling were based on these determinations.

Three papers, describing the analysis of the EXO-200 background, as derived from data, have been published~\cite{exo-200_cosmics_2016,exo-200_background_2008, Albert:2015nta}.
The key traits of that analysis methodology are reviewed here as they form the basis for nEXO's sensitivity calculations.

The event-reconstruction capability of EXO-200 is utilized to categorize events into single-site (SS) and multi-site (MS) classes. The former is predominantly composed of  $\beta$-induced signal-like events, the latter of $\gamma$-ray induced background-like  events. Point-like $\alpha$-decay induced events are identified by their large scintillation to ionization signal ratio. These separations are analyzed and determined event-by-event. The EXO-200 data analysis utilizes all event sets by performing a simultaneous fit of both the SS and MS event distributions. 
The approach of this coupled fit method offers the advantage that signal and background can be determined simultaneously. 
The energy resolution, which allows to resolve multiple peaks within the decay series, provides important constraints for the background model. 
Further signal and background discrimination through a statistical method, is achieved by utilizing the event location. On average, $\gamma$-ray  interactions occur preferentially near the detector surface. A distance to surface parameter is defined as the distance to the nearest detector surface (excluding the central cathode) and used as a third independent fit variable.
This additional analysis dimension helps refine the background model fit, which is dominated by $\gamma$-ray components. 

$^{60}$Co, $^{226}$Ra, and $^{228}$Th radioactive source  calibration data is used to constrain a \textsc{Geant4} Monte Carlo detector model. 
Differences between the model and source calibration data are utilized to quantify systematic  uncertainties. 

The tuned Monte Carlo model is used to compute probability density functions (PDFs) for all significant background contributors from all major detector components.
The set of background PDFs are fit to the data, with their normalizations free floating. The resulting PDF normalizations determine the partial contributions to the overall detector background and to the energy range where the  $0\nu\beta\beta$-peak is expected. The entire spectral, event multiplicity, and spatial information is used to determine the background for the $0\nu\beta\beta$-search. The latter is performed by means of a likelihood-ratio analysis, comparing models with and without a $0\nu\beta\beta$-peak present in the likelihood function.

An improved version of the EXO-200 analysis is described in Ref.~\cite{Albert:2017owj}, where additional information about the size of SS events are used to further discriminate between \0 decays and $\gamma$-ray scatterings. This analysis resulted in $\sim$15\% sensitivity improvement over the analysis described above. Since the approach employed by EXO-200 requires detailed understanding of the electronics response, in estimating nEXO's sensitivity a similar improvement is achieved by enhancing the event localization resolution that determines the fraction of SS events, here called SS fractions. As explained in Sec.~\ref{sec:simulation}, SS fractions in nEXO were found to be two times better than that of EXO-200, and consistent with the projected readout scheme for nEXO.

A number of important conclusions follow from the EXO-200 fit results~\cite{Albert:2015nta}, and are relevant for nEXO's design and sensitivity. The composition of the EXO-200 background in particular provides  guidance for nEXO planning.

The first important conclusion stems from the recognition that 23\% of the EXO-200 background is due to an in-situ cosmogenic radioactive isotope: \ce{^137Xe} ($\rm Q_{\beta}=4173\; keV$). \ce{^137Xe} is a background present throughout the LXe volume and thus is not reduced by a large monolithic detector volume. A deeper experimental location than WIPP is instead required for nEXO.

The contribution of cosmic rays to the EXO-200 background is described in Ref.~\cite{exo-200_cosmics_2016}. $^{137}$Xe, created by neutron capture, is the only significant cosmogenic contributor to the energy region near \Q. 
Because of its relatively long half-life of 3.8 min and high production rate, \ce{^137Xe} decay cannot be sufficiently suppressed by the active veto detector. 
Detailed calculations of several other radionuclides that are cosmogenically produced, including those due to reactions on copper, yielded background contributions of less than 1\% of that of $^{137}$Xe. 

Second, it is important to recognize that the EXO-200 background event rate is dominated by naturally-occurring radioactive impurities present in the external components.  
The noncosmogenic background originating from the xenon itself is found to be negligible.
These observations have important consequences for nEXO: the large amount of xenon is best used in a homogeneous detector, making optimal use of its $\gamma$-ray detection capability and resulting in drastically different signal and background ratio for different depths in the detector. Such ratios are calculated for nEXO based on the modeling of $\gamma$-ray attenuation, which is well understood in the few MeV energy range and routinely implemented in modern radiation transport simulation packages.

Finally, the EXO-200 data show the presence of \ce{^222Rn} in the LXe which results in background from the decay of \ce{^214Bi}. A steady-state population of $\sim$200 \ce{^222Rn} atoms was measured \cite{Albert:2015nta}, likely arising from emanation from materials in the external xenon piping system. Only 17\% of the \ce{^214Bi} daughters of \ce{^222Rn} decay in the LXe active volume, with the remaining 83\% occurring on the cathode after the \ce{^222Rn} daughter ions have drifted there. The majority of these decays are tagged using the \ce{^214Bi}-\ce{^214Po} decay. In EXO-200, \ce{^222Rn} decays in the LXe outside of the active TPC volume cannot be tagged and give rise to a small background contribution. 

The background model can further be used to test whether the pre-data taking radioassay of components, yield event rate estimations compatible with observation.
The pre-data background rate predictions agreed, within the estimated uncertainties, with the rates derived from the final fit to the low-background data \cite{Albert:2015nta},
thus indicating that the radioassay data has predictive power when
coupled with an appropriate Monte Carlo model. It is interesting to note that
the EXO-200 predictions made before data taking match as well \cite{Albert:2015nta}. 
A nontrivial conclusion follows from this observation. The event rates derived from the radioassay results assume Th and U decay chain equilibrium, while the data-driven analysis does not, thus validating the equilibrium assumption. Therefore, the methods employed during the planning and construction of EXO-200 can
be considered reliable and justify their use during the design and construction of nEXO.

\section{\lowercase{n}EXO Detector Concept}
\label{sec:nexo_design}

\begin{figure*}[tbp]
\centering
\includegraphics[width=0.49\textwidth]{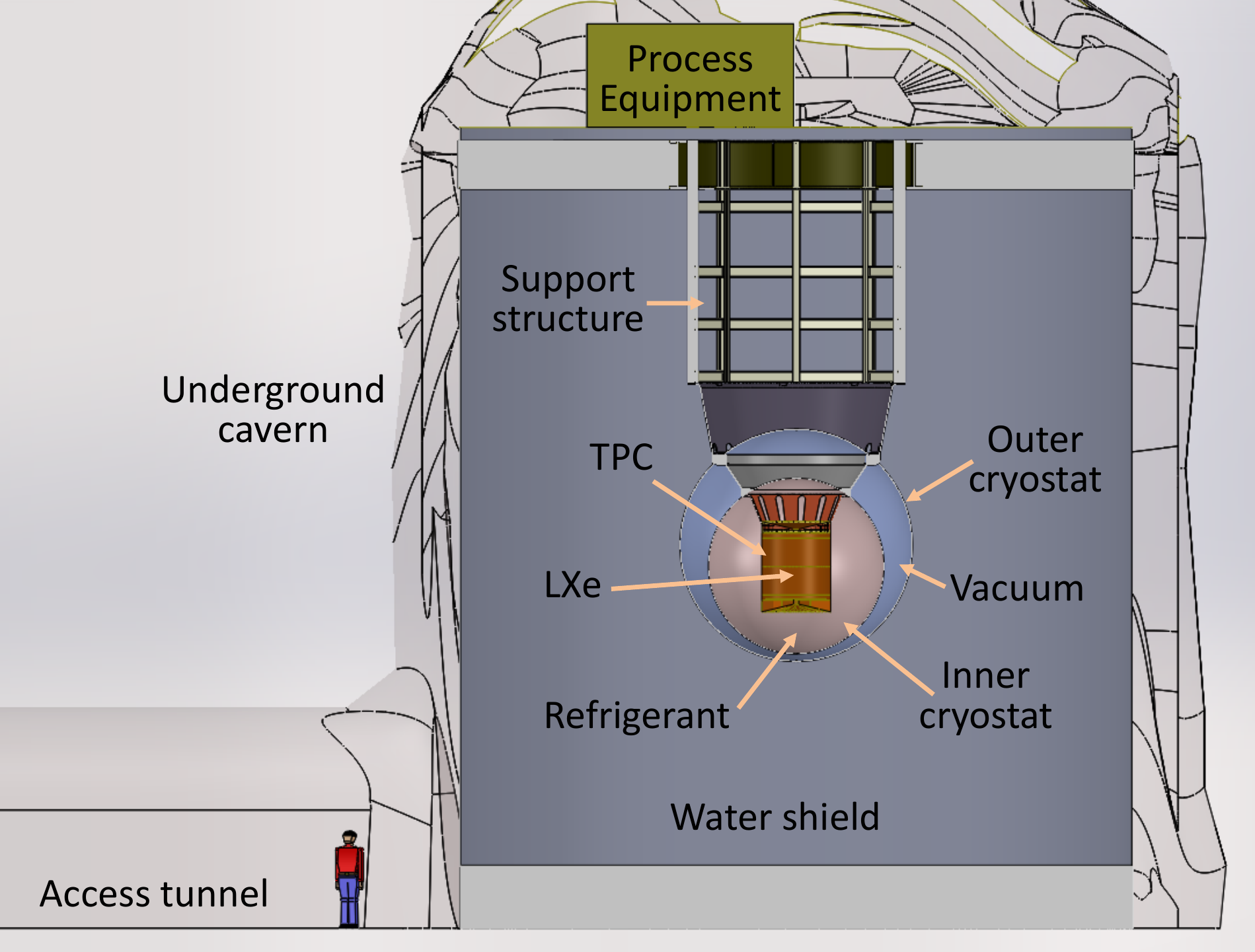}
\includegraphics[width=0.49\textwidth]{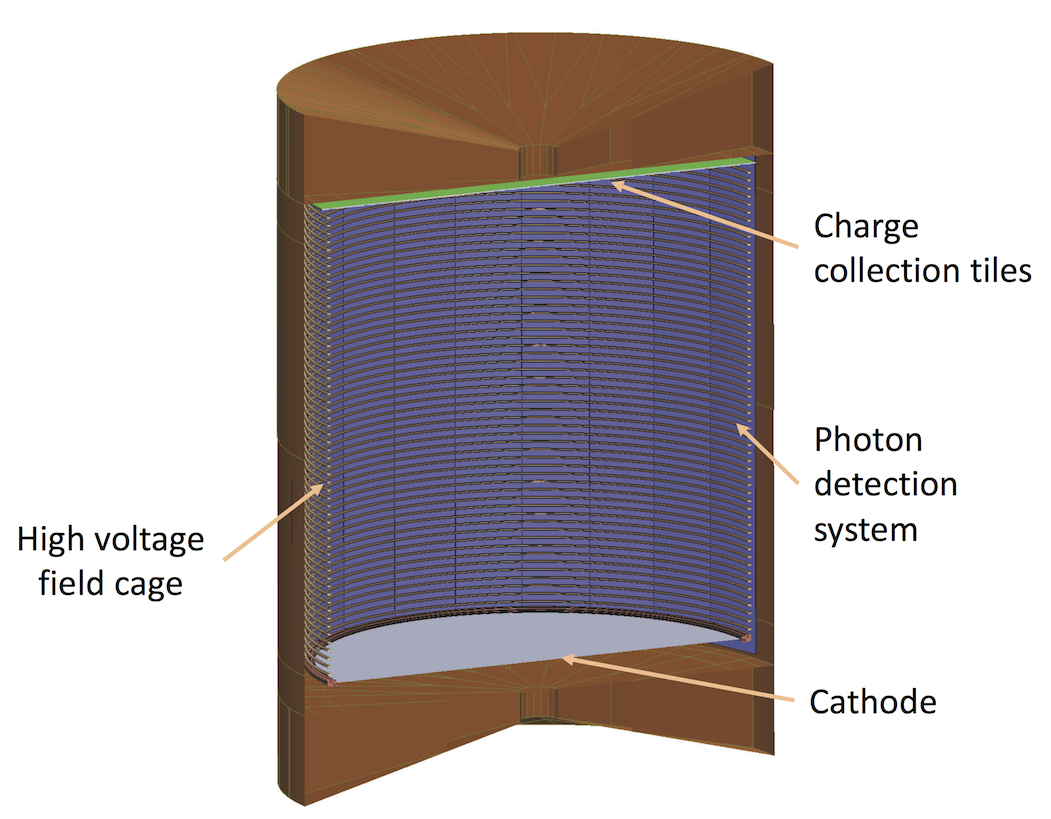}
\caption{Engineering design rendering of the nEXO experiment concept, for concreteness drawn in the SNOLAB cryopit (left). Cross-section of the TPC (right).}
\label{fig:nexo_images}
\end{figure*}

A conceptual sketch of nEXO  is shown in Fig.~\ref{fig:nexo_images}, with the  liquid xenon volume enclosed in several layers of active and passive shielding. 
The principal parameters and dimensions of the experiment are presented in Table \ref{tab:key_params}.  

\begin{table}[tbp]
\centering
\begin{tabular}{ll}
\hline
\textbf{Description} & \textbf{Value} \\ 
\hline
Liquid Xenon total mass & 5109 kg \\
TPC xenon mass & 4038 kg \\
Fiducial xenon mass & 3740 kg \\ 
\ce{^{136}Xe} enrichment level & 90\% \\ 
TPC drift height & 125.3 cm \\
TPC drift diameter & 114.8 cm \\
Drift electric field & 400 V/cm \\

TPC Vessel height & 130 cm\\
TPC Vessel diameter & 130 cm\\
HFE (Inner) Vessel diameter & 338 cm\\
Vacuum (Outer) Vessel diameter & 446 cm \\
Water Tank height & 9 m\\
Water Tank diameter & 10 m\\

\hline
\end{tabular}
\caption{Key parameters of the nEXO geometry.}
\label{tab:key_params}
\end{table}

In this concept, the TPC consists of a single homogeneous volume of 4038 kg of LXe enriched to 90\% in the candidate \0 nuclide \ce{^{136}Xe}. 
An electric field drifts electrons toward the top of the TPC. Electric-field shaping rings, connected in a chain by resistors, create a potential gradient in the LXe to achieve a uniform drift field. A drift field of 400 V/cm, similar to that of EXO-200, is planned. Deviating from the EXO-200 design, no center cathode is planned in nEXO. This is done to remove radioactivity from the center of the TPC and thereby taking full advantage of the substantial LXe-$\gamma$-ray attenuation, an important analytical tool. Hence, the design results in a single Xe-volume, delimited by cathode (at negative high voltage) and anode (at ground potential). This comes at the expense of a larger drift length and higher drift voltage requirements for equal field strength when compared to a mid-cathode design. In addition, the cryogenic charge readout electronics is only located at the top of the TPC (and the passive cathode at the bottom), thus minimizing xenon convection in the bulk of the LXe.

The field cage shaping rings are envisaged to be made of high-purity copper. Each of the 61 rings weighs 1.2 kg and are vertically spaced 2 cm apart. 
The field rings are separated by 24 1-cm high cylindrical sapphire vertical spacers (3.17 mm radial thickness) spread evenly along the ring circumference. The entire structure including the cathode and anode is held under compression by spring-tensioned sapphire rods with 1.59 mm radius that run the full length of the field cage, passing through holes in the rings and each of the spacers.

Several concepts are being investigated for the design and fabrication of the TPC cathode. For the purposes of the sensitivity calculations, the cathode is assumed to be a 0.25 mm thin copper disk held between two halves of an enlarged copper ring.  The cathode sits 30 mm above the bottom of the TPC vessel. 

Electrostatic simulations using COMSOL \cite{comsol:2017} were used to estimate the uniformity of the electric field in the drift volume. Near the edge of the field cage, distortions of the electric field lines prevent full charge collection for events within $\sim$5-mm radial distance from the rings, radially defining the boundary of the fiducial volume. Electrostatic simulations are also used to position the HV components in a way such that the maximum electric field does not exceed a surface field of 50kV/cm.  This value was derived from COMSOL simulations of EXO-200 under high-voltage configuration that provided stable operating conditions. 

An array of UV-sensitive silicon photomultipliers (SiPMs) detects  scintillation light from particle interactions in LXe.  
The SiPMs are arranged in a ``barrel'' configuration on the mantle of the cylinder inside the TPC vessel and behind the field-shaping rings.  There is no light collection at the top and bottom of the TPC volume, as the barrel region affords a better coverage and the top and bottom bases of the cylinder are  occupied by the (opaque) charge collection tiles and cathode, respectively.  
Initial work on characterizing UV-sensitive SiPMs for use in nEXO has been published in Ref.~\cite{Ostrovskiy:2015oja}.

The charge signal is envisioned to be read out on the anode by arrays of crossed strips, deposited on $10\times 10$ cm$^2$ dielectric tiles \cite{Jewell:2017dzi}. The current choice for the channel pitch is 3 mm. Studies of the optimal value are ongoing.
Compared to the wire readout used in EXO-200, the tile design simplifies construction and assembly. It further allows for the convenient placement of readout ASIC chips on the reverse of each tile.
The signal induced by the drifting electrons on each  strip is digitized and recorded independently. 

The SiPMs and their associated read-out electronics are mounted on fused-silica backing structures supported by copper components. A similar arrangement is used for the charge read-out tiles and  associated in-LXe front-end read-out electronics. 

Flexible kapton cables, similar to those used in EXO-200, bring the ionization and scintillation signals out to the external data acquisition system. In nEXO, unlike EXO-200, the front end electronics and digitization are not external to the cryostat, but rather are in the LXe in close proximity to the TPC.   In this design, the resulting relatively short cable runs for the analog signals will improve noise performance, and the reduced post-digitization cable plant reduces overall cable mass, at the cost of radiopurity constraints placed on the electronics.  In this approach, the amount of cables does not scale with the size of the detector, owing to the digitization and multiplexing possible in this configuration. 

The liquid xenon and TPC instrumentation are contained within a cylindrical vessel made of low-radioactivity copper. Initial engineering evaluations estimates the structural requirements of this vessel, including the mass of the copper, leading to a 3~mm-thick barrel with stiffened end-plates to manage static pressure loading.

Similar to EXO-200, the TPC vessel is conceived to be surrounded by a buffer of 
$\sim$33,000 kg of HFE-7000, which is critical to the cryogenic setup as a thermal bath for the LXe. The HFE buffer also provides 76 cm of shielding between the double-walled cryostat and the copper TPC vessel at their closest distance. The choice of the HFE-7000 thickness has been tuned to ensure backgrounds from the outer vessels and HFE-7000 itself are sub-dominant, as shown later.

The inner cryostat vessel, containing the HFE-7000, is modeled as spherical with a diameter of 338 cm. An outer vessel (diameter of 446 cm) provides the vacuum insulation required to maintain the inner cryostat at cryogenic temperature. The current concept employs carbon fiber  with a titanium liner as construction material for the inner and outer vessels, in place of a more standard stainless steel solution. Carbon fiber provides better mechanical properties with potentially equal or better radiopurity. Unlike stainless steel, carbon fiber also allows for on-site fabrication in case large components cannot be brought into the underground facility. Support structures within each vessel are used to fix their relative positions. The outer vessel itself will be mounted from a top platform above a water tank.

A cylindrical stainless steel tank of 10 m height and 9 m diameter and filled with water is used as active muon veto and shields against natural radioactivity in the surrounding cavern walls. As a passive neutron shield, it reduces the neutron-induced background.

The experiment's location is still under evaluation and multiple options are being investigated.
For concreteness, in the simulation the experiment is assumed to be located deep underground in the existing Cryopit at the Subdury Neutrino Observatory Laboratory (SNOLAB), near Sudbury, Ontario, Canada. With an overburden of 6010 m water equivalent (m.w.e.) \cite{snolab-manual}, this is a significant increase in depth over EXO-200 with its overburden of $1624^{+22}_{-21}$ m.w.e. \cite{exo-200_cosmics_2016},  and provides a valuable reduction in all types of cosmic-ray induced backgrounds.

\section{Backgrounds}
\label{sec:backgrounds}

\subsection{Background Sources}
\label{sec:nuclide_choice}
Detector background is a key aspect of any double-$\beta$-decay experiment. 
Building a background model involves the pre-selection of processes of interest. 
In deciding the composition of nEXO's background model, the following selection criteria were used:
\begin{enumerate}
\item The decay must release sufficient energy to interfere with the detection of the \ce{^136Xe} \0 mode.
\item The decaying nuclide must have sufficiently long half-life or be produced in steady-state in the detector. Otherwise, radioactive decay quickly diminishes the impact on the background of nuclides with a half-life time less than half a year. 
\end{enumerate}

A list of background sources that were considered during the assembly of nEXO's background model is provided in Table~\ref{tab:bkg_list}. To show the completeness of our approach, these potential background components are discussed in detail in the remainder of this section and their background contribution estimated. This provides confidence that no contribution is discarded without explicit quantifiable justification. Components resulting in background event rates $\leq0.02$ SS events/(FWHM$\cdot$yr) in the inner 2000 kg LXe volume are not further considered. The choice of this volume will be discussed later in Sec.~\ref{sec:sub_sens_results}. Components contributing above this level are then simulated in detail as described in Sec.~\ref{sec:simulation}. 

\begin{table*}[tbp]
\centering
\begin{tabular}{ll}
\hline
Background source &  \\
\hline
Long-lived radionuclides ($\gamma$- and $\beta$-emitters) in detector materials & In model \\
Th and U in water shield and laboratory rock & Negligible \\
Surface radioactivity & Negligible \\
$\alpha$ radioactivity & Negligible \\
Aboveground cosmogenic activation products & Negligible \\
Underground cosmogenic activation products in LXe & In model \\
Underground cosmogenic activation products in other detector materials & Negligible \\
\ce{^136Xe} \2 & In model \\
Activation products from ($\alpha$,n) reactions & Negligible \\
Electron-neutrino elastic scattering & Negligible \\
Neutrino capture on \ce{^136Xe} & Negligible \\
\ce{^222Rn} steady-state presence in LXe & In model \\
\ce{^222Rn} steady-state presence in water shield & Negligible \\
\hline
\end{tabular}
\caption{List of background sources that were considered during the development of nEXO's background model and whether they were included in the sensitivity calculations. Details are provided in the text. }
\label{tab:bkg_list}
\end{table*}

\subsubsection*{Long-lived Radionuclides}
The naturally occurring radioactivity of the \ce{^232Th} and \ce{^238U} nuclides fulfills the decay time selection criteria and, as a result, the decay chain daughters from each nuclide are included in the background model (as noted in Table~\ref{tab:isotopes_generated}). Of particular interest is the \ce{^238U} daughter \ce{^214Bi} which decay includes a $\gamma$-ray line at 2448 keV.

Long-lived nuclides such as $^{137}$Cs, $^{60}$Co, and \ce{^40K} are also considered. While they do not contribute to the \0 background due to energy and $\gamma$-ray multiplicity, they significantly affect the measurement of the \2 decay and are therefore systematically tracked as part of nEXO's materials analysis program. \ce{^40K} was explicitly included in the model as representative of low-energy spectral features. The probability that a $^{60}$Co decay result in a SS event with energy within \Q$\pm$FWHM/2 was estimated from Monte Carlo to be negligible ($<2.3\times10^{-8}$ at 90\% C.L. in the fiducial volume of 3740 kg). \ce{^60Co}-induced background is therefore negligible. The \2 decay of \ce{^136Xe} is included in the model.
\ce{^26Al} is not currently included in the background model, but is planned for future study. 

Only $\gamma$ and $\beta$ decays from the nuclides above directly create background events in nEXO. Decays emitting only $\alpha$ particles are rejected with high efficiency using charge/light ratio analysis \cite{Albert:2015vma}. Secondary radionuclide production, e.g., through ($\alpha$,n) reactions, is discussed below.

Background radioactivity was then further subdivided into bulk and surface activities.
The materials analysis program described in Sec.~\ref{sec:radioassay} tests all materials of interest for their bulk radioactivity content. It is assumed that surface activities can be mitigated by an appropriate surface treatment, cleaning, clean machining, and/or etching. This strategy was effective in EXO-200.
The list of bulk radionuclides included in the Monte Carlo for each detector component in the nEXO background model is given in Table~\ref{tab:components}.

Several studies have demonstrated that a set of long-lived radionuclides from certain components of the experiment do not contribute significantly. For example,
radiopurity levels of \ce{^238U} and \ce{^232Th} equal or better than 1 ppt have been achieved in water \cite{Boger:1999bb}. A dedicated simulation of these impurities in the water shield showed that the \ce{^208Tl} and \ce{^214Bi} nuclides present in the decay chains contribute about 0.015 SS events/(FWHM$\cdot$yr) in the LXe volume inside the TPC field cage. The contribution of radioactivity in the shielding water was thus not considered in the background model.

Another source shown to be negligible is the natural radioactivity from the  walls of the underground laboratory which was studied using a simplified experimental geometry. 
A layer of concrete and shotcrete covers SNOLAB's cryopit rock walls. \ce{^{226}Ra}(\ce{^238U}) and \ce{^232Th} in these materials at specific activities from \cite{SNOLABRock2007} 
were found to contribute only 0.04 SS events/(FWHM$\cdot$yr) in the LXe volume inside the TPC field cage. Increasing the water shield radius by 1 m, which e.g., still fits comfortably in the SNOLAB cryopit, would further reduce this background by a factor of $\sim$100. 

\subsubsection*{Cosmogenically-created Radionuclides}
Radionuclides with a half-life of less than 0.5 year were considered if they can be created by the interaction of the cosmic radiation with a material of interest.  
The estimation of this background class had two components: activation while materials are stored, handled, or machined above ground and the steady-state production underground. The former results in guidelines for the exposure management, the latter defines the requirements for the overburden. 

Above-ground radio-nuclide production is important for all passive detector materials. In particular, the production of radio-nuclides in copper (e.g., $^{56}$Co and $^{60}$Co) was estimated and found to be acceptable with proper management of the cosmic ray exposure. 
Because the xenon will be continuously purified during detector operation, long-lived spallation products created by the cosmic radiation while the xenon is above ground (e.g., $^{137}$Cs) are not a concern. Xenon has no long-lived cosmogenically-produced isotope.  

EXO-200 data was used to quantify a broad range of cosmogenic backgrounds~\cite{exo-200_cosmics_2016} that would arise during underground operation. This was accomplished by testing GEANT4 and FLUKA Monte Carlo simulations against data, thus validating the models. These simulations were used, appropriately modified, to estimate cosmogenic backgrounds in nEXO, providing more confidence in the procedure. As a result, with sufficient overburden such as that available, e.g., at SNOLAB, all cosmogenic backgrounds except $^{137}$Xe are negligible. $^{137}$Xe, which $\beta$ decays with a $Q$ value of 4173 keV, is therefore the only cosmogenic activity contained in the nEXO background model. 

At SNOLAB, the steady-state production of cosmogenic \ce{^137Xe} in nEXO was estimated using FLUKA at $2.2\times 10^{-3}$  atoms/(kg$\cdot$yr). Siting nEXO at locations with similar overburden, like China Jinping Underground Laboratory \cite{Chen:2012fm}, would also result in an acceptable cosmogenic nuclide production, even without active vetoing. 
The depth of Laboratori Nazionali del Gran Sasso (LNGS) \cite{Votano:2012fr}, Italy, was found to be marginally acceptable and require the development of a more sophisticated active veto system. 
The Sanford Underground Research Facility (SURF), South Dakota, USA, \cite{Lesko:2011qk,Lesko:2012fp} at 4850-feet depth is adequate assuming a simple active veto scheme.

\subsubsection*{Neutrino-induced Backgrounds}
Interactions of solar neutrinos in the detector are a potential source of background for \0 experiments, as discussed in Refs.~\cite{Vogel-note2005,Ejiri:2013jda,deBarros:2011qq}.

Electron-neutrino elastic scattering ($\nu + e^- \rightarrow \nu + e^-$) in the detector  volume results in the emission of energetic electrons that can mimic the signature of a \0 event. Using the background rate for this reaction in \ce{^136Xe} from \cite{deBarros:2011qq},  
$\sim$0.02 SS events/(FWHM$\cdot$yr) are expected in nEXO's inner 2000 kg of LXe. 
At this level this background is expected to be small compared to other backgrounds and thus we have chosen to set this as the level at which backgrounds are excluded from further investigation for the sensitivity calculation presented in this work.

The neutrino capture process via the charged-current reaction $\nu+$\ce{^136Xe}$\rightarrow e^-+$\ce{^136Cs} also contributes background events due to (1) the prompt $e^-$ combined with any $\gamma$-ray emitted from the \ce{^136Cs} de-excitation, and (2) the delayed decay of \ce{^136Cs} into \ce{^136Ba} with a half-life of 13.16 days and $Q=2548.2$ keV.

The rate of events due directly to the neutrino capture process and falling near \Q is expected to be very small. The dominant \ce{^7Be} and other low energy solar neutrinos cannot produce enough visible energy to reach \Q while almost all of the events due to the \ce{^8B} flux give too much visible energy. Estimates of the true background rate have been made following Refs.~\cite{Vogel-note2012,Ejiri:2013jda,Elliott:2017bui}, predicting backgrounds rates totally negligible for nEXO.

The total charged-current interaction rate for solar neutrinos on \ce{^136Xe} has been shown \cite{Vogel-note2012,Ejiri:2013jda,Elliott:2017bui} to be about  20 interactions/(2000 kg$\cdot$yr). In Ref.~\cite{Ejiri:2013jda} it is estimated that the rate of events within \Q$\pm$FWHM/2 would be about 0.6 events/(2000 kg$\cdot$yr) and a similar results is found in Ref.~\cite{Vogel-note2012}. The decay proceeds with typically three gamma rays in cascade together with the electron with a total energy release of 2548 keV so very high single site rejection is expected. Indeed, a Monte Carlo simulation of $10^{7}$ decays of \ce{^136Cs} in nEXO's LXe volume gave no events that satisfied both the energy and the single site criterion.
Even assuming that \ce{^136Cs} is not removed by the LXe purification system or does not freeze-out on metal surfaces, the  decay of all \ce{^136Cs} in the LXe would thus result in a negligible background rate in nEXO. 

Finally, solar neutrinos can interact with \ce{^136Xe} through an inelastic scattering neutral-current interaction. Such a process could excite a $1^+$ state which would have a high branching rate back to the ground state, thus potentially resulting in SS event. We are not aware of any $1^+$ state with energy near \Q~and therefore neglected this background source.

\subsubsection*{Radionuclides from ($\alpha$,n) Reactions}
Deposition of the $\alpha$-unstable $^{222}$Rn daughter $^{210}$Po  can create background through ($\alpha$,n) reactions with low-Z detector materials such as F, C, O, Al, and Si which are contained in nEXO's HFE-7000, sapphire, and quartz. The emitted neutrons can subsequently produce \ce{^137Xe} when capturing on \ce{^136Xe}.

A calculation was performed to determine the allowable exposure time of these materials to standard laboratory air (25 Bq/m$3$ \ce{^222Rn}) before radon daughter plate-out results in more than 0.01 events/(FWHM$\cdot$yr$\cdot$3000 kg) of background.
Neutron yields for the relevant ($\alpha$,n) reactions  were calculated from tabulated stopping power and cross-section data, and used as input into a FLUKA simulation to determine the position distribution and probability of neutron captures on \ce{^136Xe} in nEXO from neutrons generated on the surface of the relevant components. Capture events were found to be uniformly distributed in the LXe. The nEXO \textsc{Geant4} Monte Carlo provided the fraction of \ce{^137Xe} decays that result in a SS energy deposition within \Q$\pm$FWHM/2.  
Measured radon-daughter deposition rates \cite{Guiseppe:2011mj} were used. A constant \ce{^210Po} decay rate has been assumed, leading to an over estimate of the neutron production as the slow Po-growth is not accounted for.

Of the surfaces considered, those in contact with the HFE-7000 would be subject to the most restrictive exposure time constraint of 7.5 yr/m$^2$. Therefore, the steady-state production of \ce{^137Xe} from $\alpha$-induced neutrons can be neglected under the assumption that proper surface treatment and handling will be performed during construction.

\subsubsection*{\ce{^222Rn}}
Contributions to the background rate from the \ce{^222Rn} daughter \ce{^214Bi} must be considered because \ce{^214Bi} emits a $\gamma$-ray with an energy only 10 keV lower than the \Q value. As a result, any process that contributes a steady-state population of \ce{^222Rn} inside the LXe volume is important. For the purposes of estimating the sensitivity, 
it is assumed that nEXO will have 600 \ce{^222Rn} atoms continuously present in the LXe, a factor of 3 higher than observed in EXO-200. This factor is based on an estimate of the expected inner surface area of the xenon recirculation system in nEXO relative to EXO-200.

Events from the \ce{^222Rn} decay chain can be tagged using a Bi-Po veto which identifies time- and space-correlated $\beta$ and $\alpha$ decays.  An equivalent efficiency as that achieved by EXO-200 was assumed for nEXO to reject Bi-Po events from \ce{^214Bi} decaying directly in the LXe volume inside the TPC field cage. 
Tagging or vetoing of \ce{^214Bi} decays in the LXe outside the TPC field cage may be possible in nEXO by exploiting the light collected by the SiPMs on the barrel (not possible in EXO-200). 
The ability to identify \ce{^214Bi} decays from \ce{^222Rn} daughters that have drifted on to the cathode is the subject of ongoing studies, including special cathode designs and analysis techniques. 
The background model presented in this work combines \ce{^214Bi} in the region outside the TPC field cage and on the cathode into one term, and assumes a tagging efficiency of $\sim$40\% for these decays. 

\ce{^222Rn} can dissolve into the water of the water shield, producing a background from $\gamma$-rays penetrating in the TPC. Assuming a Rn concentration level of 9$\times10^{-9}$ Bq/kg total from all components exposed to water (steel tank, pumps, other materials, etc.), Monte Carlo simulations show the background contribution to be negligible. This concentration requirement is satisfied if the wetted steel has no more than 10 ppb U, expected to be easily achievable.

\subsection{Detector Component Materials and Assays}
\label{sec:radioassay}

Radioassay of detector materials is a central aspect of demonstrating feasibility and from which to base  sensitivity projections on realistic assumptions. 
This activity is intimately connected with the Monte Carlo and engineering efforts, enabling the preliminary detector engineering to be performed with appropriate low activity materials. This is essential for the iterative development of a compelling and realistic experiment concept.
nEXO's approach follows the example of the successful EXO-200 materials certification and background estimation efforts~\cite{exo-200_background_2008,Leonard:2017okt}.

This effort is being conducted at multiple collaborating nEXO labs worldwide. In addition, commercial services are utilized whenever appropriate. The following assay techniques have been utilized (benchmark sensitivities given in square brackets, for Th/U in techniques 1--4):
\begin{enumerate}
\item Aboveground and underground low background $\gamma$-ray spectrometry. Purpose designed 
shielded Ge detectors are utilized. [routine: 200/35 ppt, 2.3/1.2 ppt achieved with very large samples]\label{meth:ge}
\item Inductively coupled plasma mass spectrometry (ICPMS). [routine:  1/1 ppt, 0.008/0.01 ppt
achieved with pre-concentration]\cite{LaFerriere:2014rva}\label{meth:icpms}
\item Glow discharge plasma mass spectroscopy (GDMS). [10/10 ppt]\label{meth:gdms}
\item Neutron activation analysis (NAA). [routine: 1/1 ppt, 0.02/0.02 ppt achieved with sample pre-concentration]\label{meth:naa}
\item Radon out-gassing via electrostatically-boosted solid state detection or
liquid scintillation counting [3 atoms/(m$^2$ d)]\label{meth:rn}
\item Low background $\alpha$-counting using Si solid state detector [\ce{^210Po}: 30 mBq/m$^2$]\label{meth:a}
\end{enumerate}

These methods complement each other.
ICPMS and NAA offer the best sensitivity at the levels required for the innermost and most demanding detector components.
However, converting the concentrations of nuclides at the head of the decay chains, determined by these methods, into background rates requires assumptions regarding the chain equilibrium. Assuming secular equilibrium is realized, these techniques  can only estimate, rather than rigorously predict, the expected background rates. 
$\gamma$ spectroscopy with Ge detectors directly determines the background relevant Th and U-chain members $^{208}$Tl and $^{214}$Bi. $\gamma$-ray spectroscopy is further used to probe for short lived activities (cosmogenic or man-made). 
Radon counting directly probes backgrounds from this nuclide.

Table~\ref{tab:rad} lists the detector materials and activities entering into the current background estimate. Except for three cited entries, Table~\ref{tab:rad} presents new nEXO measurements of these materials, not previously published elsewhere.

The nEXO material assay data are recorded and stored in an online database. The data catalog systematically keeps track of measurement values and uncertainties to
enable the computation of a total background rate, accounting for errors. For convenience Table~\ref{tab:rad}
lists 90\% C.L. limits whenever the value is consistent with zero at 90\% C.L., assuming the errors follow
the normal distribution. The limit conversion uses the ``flip-flop'' method \cite{Feldman:1997qc} to avoid inflating the sensitivity for un-physical negative concentrations. For measurements near the limit of sensitivity, which is often the case for nEXO samples, seemingly un-physical results are  encountered and dealt with in a consistent fashion, as described above.

\begin{table*}[tbp]
\centering
\begin{tabular}{lllccccc}
\hline
\textbf{Material}     & \textbf{Supplier}               & \textbf{Method}  & \textbf{K   }     & \textbf{Th }         & \textbf{U }  & \textbf{\ce{^60Co} }     \\ 
     &               &   & \textbf{[ppb]  }     & \textbf{ [ppt]}         & \textbf{U [ppt]}  & \textbf{[$\mu$Bq/kg]}     \\ \hline
Copper       & Aurubis                & ICPMS/Ge/GDMS   & $<$0.7   & 0.13$\pm$0.06    & 0.26$\pm$0.01  & $<$3.2  \\
Sapphire     & GTAT                   & NAA     & 9.5$\pm$2.0   & 6.0$\pm$1.0      & $<$8.9      & -     \\
Quartz       & Heraeus                & NAA     & 0.55$\pm$0.04 & $<$0.23          & $<$1.5       & - \\
SiPM         & FBK                    & ICPMS/NAA   & $<$8.7        & 0.45$\pm$0.12    & 0.86$\pm$0.05  & -  \\
Epoxy$^*$        & Epoxies Etc.           & NAA     & $<$20         & $<$23            & $<$44    & -      \\
Kapton$^*$   & Nippon Steel Cables& ICPMS   & -             & $<$2.3 pg/cm$^2$ & 4.7$\pm$0.7 pg/cm$^2$ & - \\
HFE$^*$      & 3M HFE-7000            & NAA     & $<$0.6        & $<$0.015         & $<$0.015    &  - \\
Carbon Fiber & Mitsubishi Grafil      & Ge      & 550$\pm$51    & 58$\pm$19        & 19$\pm$8  & -  \\
ASICs & BNL & ICPMS & - & 25.7$\pm$0.7 & 13.2$\pm$0.1 \\
Titanium & TIMET & Ge & $<$3.3 & 57$\pm$5 & $<$7.3 & - \\
Water        & SNOLAB                 & Assumed & $<$1000       & $<$1             & $<$1   &-    \\ \hline
\end{tabular}
\caption{Materials, analysis method and radioactivity concentrations entering the nEXO background model. Data for entries marked with a $^*$ were taken from the EXO-200 materials certification program. Data for titanium are from Table~VI of Ref.~\cite{Akerib:2017iwt}. Limits are
stated at 90\% C.L. and were computed using the ``flip-flop'' method \cite{Feldman:1997qc}.}
\label{tab:rad}
\end{table*}

\section{nEXO Simulations} \label{sec:simulation}

\subsection{Geometry and Event Generation} 

A \textsc{Geant4}-based application \cite{GEANT4-2003} is the primary tool used to simulate energy depositions in the detector. 
Table~\ref{tab:EMPhysics} describes the electromagnetic physics processes from \textsc{Geant4.10.02} used in this simulation. Processes and particles not listed are left to the default \textsc{Geant4} physics lists.

\begin{table}
\centering
\renewcommand{\arraystretch}{1.25}
\begin{tabular}{lll}
\hline
\textbf{Particle} & \textbf{Process} & \textbf{Energy range}
\\ \hline
$\gamma$          & Livermore EM     & $<  1~\si{\giga\electronvolt}$  \\
$e^{-},e^{+}$     & Urb\'an Multiple Scattering     & $<  100~\si{\mega\electronvolt}$ \\
                  & Wentzel-VI Multiple Scattering  & $>  100~\si{\mega\electronvolt}$ \\
                  & Coulomb Single Scattering     & $>  100~\si{\mega\electronvolt}$ \\
$e^{-}$           & Livermore Ionization    	 & $<  100~\si{\kilo\electronvolt}$ \\
                  & Livermore Bremsstrahlung  & $<  1~\si{\giga\electronvolt}$\\
\hline              
\end{tabular}
\caption{Electromagnetic processes used in nEXO simulations. Energy ranges and additional particles that are not specified (muons, mesons, etc.) are governed by \textsc{Geant4.10.02} default physics.}
\label{tab:EMPhysics}
\end{table}

\begin{figure*}[tbp]
\centering
\includegraphics[width=0.46\textwidth]{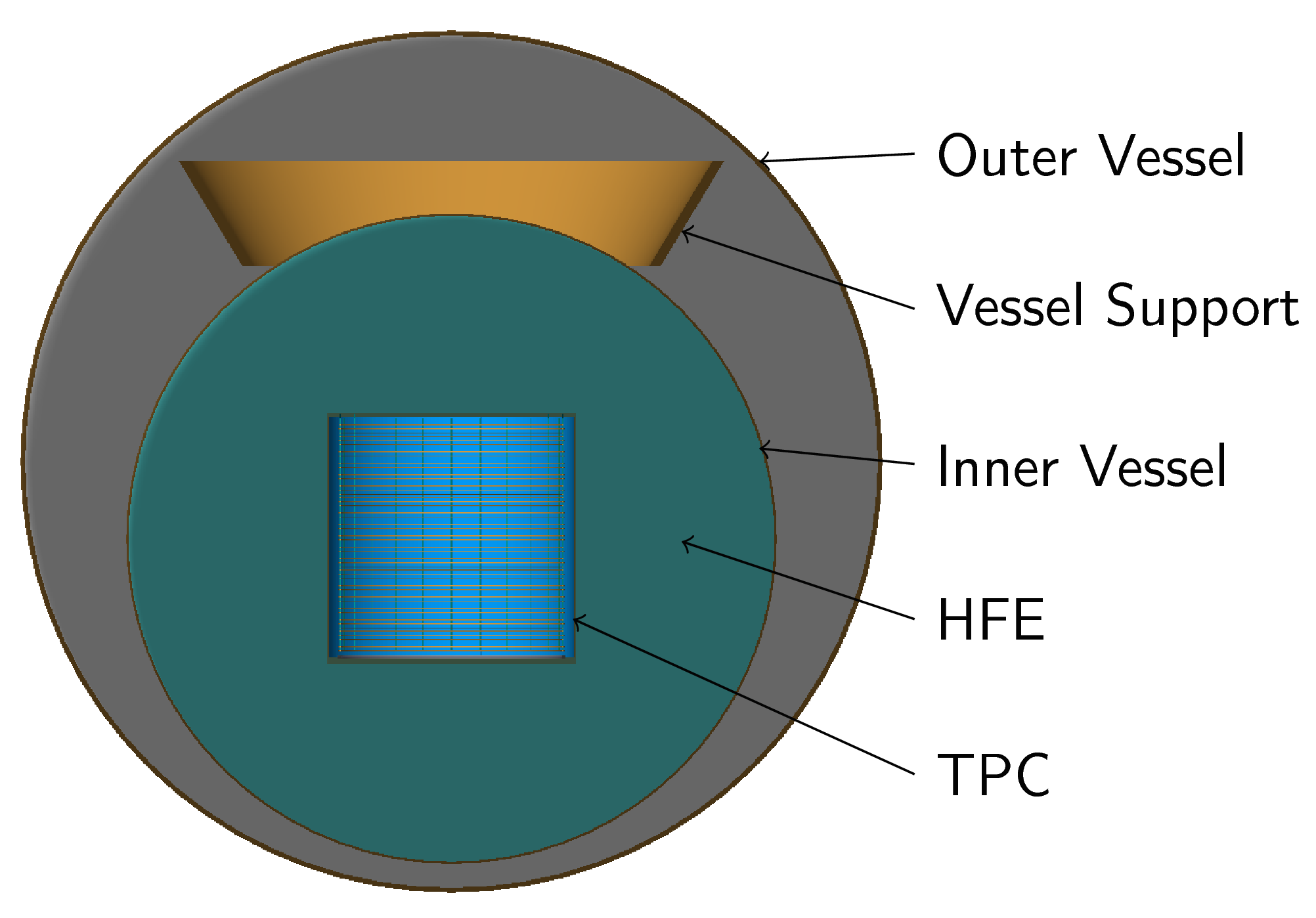}
\includegraphics[width=0.53\textwidth]{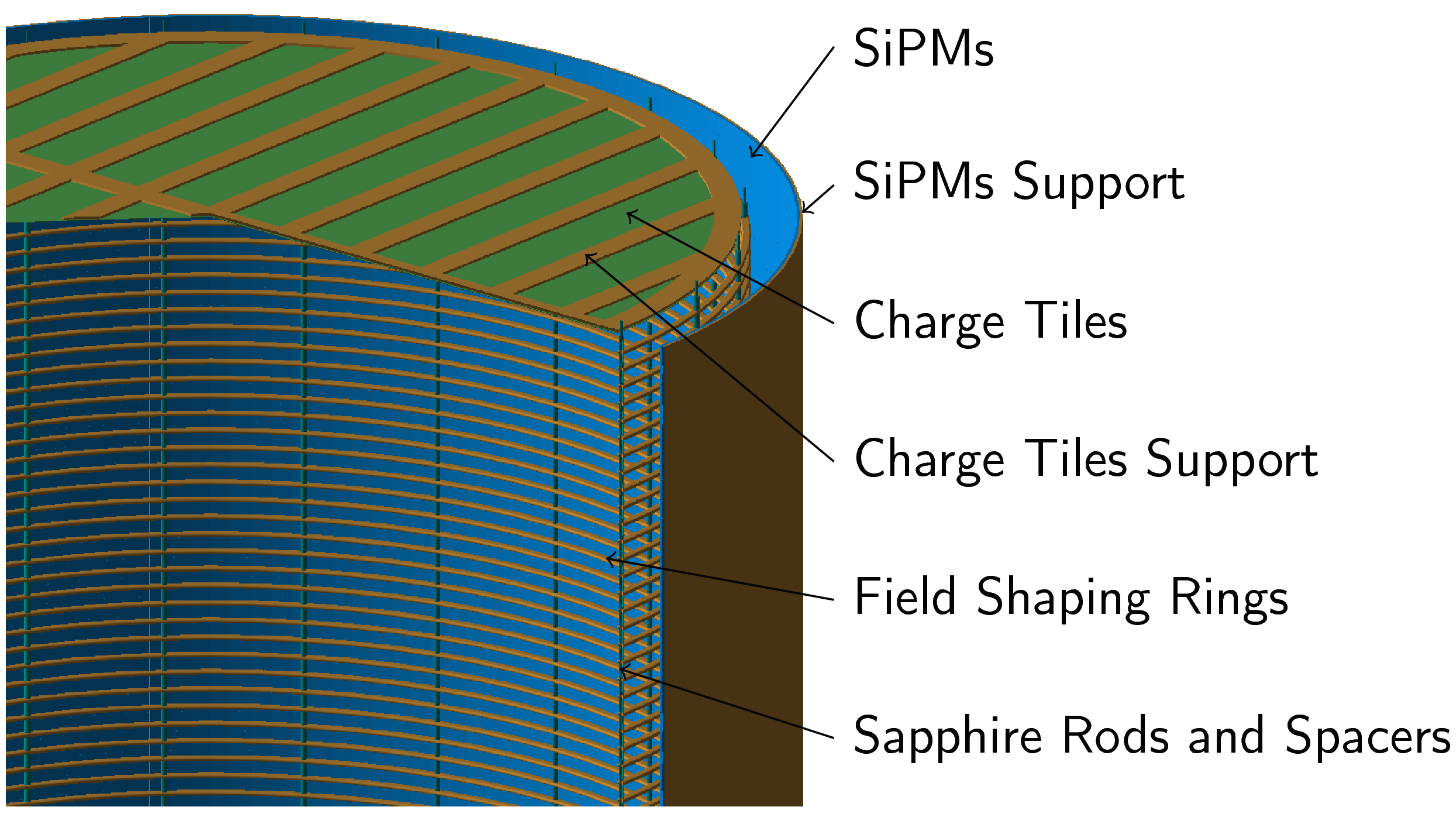}
\caption{Visualization of the \textsc{Geant4} simulation geometry. A cross-section of the components within the outer vessel are shown (left) with a close-up of the TPC (right). The underground laboratory walls and the large water shield surrounding the outer vessel are not shown but are included in the full \textsc{Geant4} model geometry. }
\label{fig:MCvisualization}
\end{figure*}

A \textsc{Geant4} geometry model implementation of the detector design described in Sec.~\ref{sec:nexo_design} was developed. The geometry uses standard \textsc{Geant4} shapes to facilitate modifications allowing evaluation of design alternatives. While approximate geometries were used, care was taken to ensure  all significant components are included, accounting for mass and materials properly.  Visualizations of the simulated geometry are shown in Fig.~\ref{fig:MCvisualization}. The list of components is provided in Table \ref{tab:components}.
Some detector components have not yet been included in the model. These are the high-voltage feed-through, the external support structures, and planned external calibration guide tubes. Due to their position and/or size, these missing components are not expected to contribute significantly to the overall background event rate and will be incorporated as the design becomes more detailed. 

\begin{table*}[tbp]
\centering
\renewcommand{\arraystretch}{1.25}
\begin{tabular}{lllc}
\hline
\textbf{Component}    & \textbf{Nuclides}                                      & \textbf{Material} & \textbf{Mass or}              \\ 
& \textbf{Simulated} & & \textbf{Surface Area}\\
\hline 
\rule{0pt}{1.2em}Outer Cryostat        & \ce{^{238}U}, \ce{^{232}Th}, \ce{^{40}K} & Carbon Fiber  &  1774 kg                                \\
Inner Cryostat        & \ce{^{238}U}, \ce{^{232}Th}, \ce{^{40}K} & Carbon Fiber  &  338 kg                                 \\
Inner Cryostat Liner  & \ce{^{238}U}, \ce{^{232}Th}              & Titanium          &  161.4 kg                             \\
HFE                   & \ce{^{238}U}, \ce{^{232}Th}                            & HFE-7000          & 32700 kg                              \\
TPC Vessel            & \ce{^{238}U}, \ce{^{232}Th}                            & Copper            & 553.4 kg                              \\
Cathode     & \ce{^{238}U}, \ce{^{232}Th}                            & Copper            & 13.02 kg                               \\
Field Rings (FR)          & \ce{^{238}U}, \ce{^{232}Th}                            & Copper            & 73.2 kg                               \\
FR Support Leg        & \ce{^{238}U}, \ce{^{232}Th}, \ce{^{40}K}               & Sapphire          & 0.94 kg               \\
FR Support Spacer     & \ce{^{238}U}, \ce{^{232}Th}, \ce{^{40}K}               & Sapphire          & 2.21 kg                                      \\
SiPM                  & \ce{^{238}U}, \ce{^{232}Th}, \ce{^{40}K}               & SiPM              & 4.69 kg                               \\
SiPM Support          & \ce{^{238}U}, \ce{^{232}Th}                            & Copper            & 136.4 kg                              \\
SiPM Module Backing   & \ce{^{238}U}, \ce{^{232}Th}                            & Quartz            & 3.2 kg                                      \\
SiPM Electronics      & \ce{^{238}U}, \ce{^{232}Th}                            & ASICs           & 2.04 kg                                      \\
SiPM Glue             & \ce{^{238}U}, \ce{^{232}Th}, \ce{^{40}K}               & Epoxy          & 0.12 kg                               \\
SiPM Cables           & \ce{^{238}U}, \ce{^{232}Th}                            & Kapton            & $1\times 10^4$ cm$^2$ \\
Charge Module Cables  & \ce{^{238}U}, \ce{^{232}Th}                            & Kapton            & $1\times 10^4$ cm$^2$                \\
Charge Module Electronics    & \ce{^{238}U}, \ce{^{232}Th}                            & ASICs           & 1.0 kg                                      \\
Charge Module Glue    & \ce{^{238}U}, \ce{^{232}Th}, \ce{^{40}K}               & Epoxy          & 0.35 kg                                      \\
Charge Module Support & \ce{^{238}U}, \ce{^{232}Th}                            & Copper            & 11.7 kg                               \\
Charge Module Backing & \ce{^{238}U}, \ce{^{232}Th}                            & Quartz            & 0.94 kg                               \\
TPC LXe Volume                   & \ce{^{137}Xe}, \ce{^{222}Rn}, \2, \0                   & Xenon             & 4038 kg                               \\
Outer LXe Volume                   & \ce{^{137}Xe}, \ce{^{222}Rn}, \2, \0                   & Xenon             & 1071 kg                               \\ \hline
\end{tabular}
\caption{List of detector components included in the \textsc{Geant4} model of nEXO with their material, nuclides simulated, and mass or surface area.}
\label{tab:components}
\end{table*}

The \textsc{Geant4} Monte Carlo is used  to generate and transport particles from radioactive decays of the relevant nuclides identified in Sec.~\ref{sec:backgrounds} (also reported in Table~\ref{tab:components}). 
To save computing time, a subset of  daughters of the \ce{^{238}U} and \ce{^{232}Th} chains are simulated independently and their resulting energy deposits subsequently merged with the appropriate branching ratios. Radionuclide selection is based on emission of $\gamma$ radiation with energy $>$100 keV and with intensity $>1$\%. Table~\ref{tab:isotopes_generated} lists those nuclides individually simulated for the \ce{^{238}U} and \ce{^{232}Th} decay chains.
Timing between the decays of daughter nuclides is not considered in the current reconstruction algorithm. 
Generation of \2 decays is performed using the algorithm in \cite{massiveneutrinos},  validated in the EXO-200 analysis. It is biased to ensure a minimum of 10$^5$ simulated events with summed electrons energy above 2250 keV.

\begin{table}[tbp]
\centering
\renewcommand{\arraystretch}{1.3}
\begin{tabular}{ll}
\hline
Decay Chain   & Nuclides generated                                                  \\ 
\hline
\ce{^238U}    & \ce{^{234}Pa},\ce{^{226}Ra}, \ce{^{214}Pb}, \ce{^{214}Bi}                 \\
\ce{^{232}Th} & \ce{^{228}Ac}, \ce{^{224}Ra}, \ce{^{212}Pb}, \ce{^{212}Bi}, \ce{^{208}Tl} \\
\hline
\end{tabular}
\caption{List of nuclides individually simulated for the \ce{^{238}U} and \ce{^{232}Th} decay chains.}
\label{tab:isotopes_generated}
\end{table}

A FLUKA \cite{FLUKA} model of nEXO has been developed for use in dedicated studies involving cosmogenic activation and neutron interactions.

\subsection{Detector Response\label{sec:MC-recon}}
The experience of EXO-200 provides a basis for estimating nEXO's detection performance, in particular for reconstructing energy, position, and multiplicity of each event.

The output of the \textsc{Geant4} simulation is reconstructed through  software that applies detector effects from charge and light transport, and performs event analysis to extract the relevant event parameters. 

First, to mimic the detector ability to identify distinct interaction sites, the \textsc{Geant4} energy deposits are aggregated by an algorithm that produces clusters of $\sim$3 mm radius based on the relative position and energy of each deposit.
This choice produces a SS fraction for background $\gamma$-ray events within \Q$\pm1.7\cdot$FHWM that are about half ($\sim$10\%) of that seen in EXO-200 ($\sim$20\%). This reflects both our estimates of the projected hardware improvements (most notably the factor 3 reduction in charge channel pitch) as well as the sensitivity improvement obtained in EXO-200 Phase II using information about the SS cluster size \cite{Albert:2017owj}. A fiducial cut removes clusters that fall within 1.5 cm of the inner edges of the TPC field cage.

nEXO's energy resolution derives from extrapolating EXO-200's demonstrated performance. Accounting for the lower noise of nEXO's SiPM read-out, coupled with an expected light collection efficiency similar to EXO-200, nEXO's energy resolution is estimated as $\sigma/\Q=1\%$. 

The collection efficiency of the scintillation light is a key factor in determining the energy resolution in LXe TPCs. nEXO's \textsc{Geant4} model was used to propagate scintillation photons through the detector until they are absorbed in the SiPMs depletion region. The optical properties assumed for the detector surfaces are listed in Table ~\ref{tab:opticalProperties}. 
Considering that a SiPM light detection efficiency of 15\% at 175 nm is achievable \cite{Ostrovskiy:2015oja}, simulations show that it is possible to obtain an average light collection efficiency of $\sim$7\%, comparable to that of EXO-200.

\begin{table}[tbp]
\centering
\begin{tabular}{lll}
\hline
Components & Properties & Value \\
\hline
Cathode & Reflectivity & 80 \% \\
Field Rings & Reflectivity & 80 \% \\
Anode (gold) & Refractive index & $1.34+0.95i$ \\
SiO2 & Refractive index & $1.61+0i$ \\
Si & Refractive index & $0.682+2.45i$ \\
LXe & Refractive index & 1.66 \\
LXe & Scattering length & 40 cm \\
LXe & Absorption length & 20 m \\
\hline
\end{tabular}
\caption{Optical properties of surfaces used in simulating nEXO's collection efficiency for xenon scintillation light. Cathode and field rings are assumed to have a reflective coating. Values are given at a wavelength of 175 nm.}
\label{tab:opticalProperties}
\end{table}

Due to the long drift length ($\sim 125$ cm), an electron lifetime $>10$ ms is deemed necessary so not to degrade the energy resolution. Electron lifetimes near this range have already been demonstrated \cite{exo-200_detector_paper,Albert:2013gpz}, and nEXO plans to limit electronegative impurities in the LXe arising from surface outgassing by utilizing substantially fewer plastic components than EXO-200 and thanks to a more favorable surface-to-volume ratio.

In the post-simulation reconstruction code, energy resolution is applied through convolution with a Gaussian distribution with a width obtained from a quadratic function analogous to that observed in EXO-200 \cite{exo200-PRL2012} and coefficients scaled to achieve the expected nEXO resolution $\sigma/\Q=1\%$. 

In addition to the event energy and distance to the nearest surface, the reconstruction also computes the event multiplicity from the number of clusters, thus classifying SS and MS events. 
The final output of the basic reconstruction consists of 2D histograms of energy vs distance to surface for each of the experiment's component and radionuclide of interest. Separate histogram for SS and MS events are created.

The reconstruction algorithm employed for this work is simple and computationally inexpensive and was shown to be able to reproduce EXO-200 spectral shapes, providing confidence in the methodology while more advanced modeling is developed and validated. 

A preliminary validation of the simple energy clustering algorithm was undertaken using a more sophisticated approach. 
Simulated energy deposits in the LXe were converted to electrons and photons using the Noble Element Simulation Technique (NEST) tool, designed to accurately model the scintillation and ionization yields in LXe and other noble-element media \cite{NEST2013}. These photons and electrons were then propagated to collection, taking into account electron diffusion (electron diffusion coefficients were taken from \cite{EXO200-electronDriftDiffusion-PhysRevC.95.025502}). Capture by electronegative impurities was neglected under the assumption of 10 ms electron lifetime and a maximum drift time of $\sim$ms at the nominal electric field. The induced signals on the charge tiles were computed using the Shockley-Ramo theorem \cite{Shockley:1938,Ramo:1939} and a COMSOL-generated weighting potential. Electronics sampling rate, noise, and threshold were applied to the waveforms which were then analyzed for signal amplitude, rise time, collection channel multiplicity, and position. A data analysis using multivariate classification techniques was performed on simulated samples of \0 (signal) and 2.5 MeV $\gamma$ (background-like) events. The resulting SS/MS discrimination performance is consistent with that of the simple clustering algorithm described above. 

Uncertainties due to systematic effects have not yet been investigated in detail.  These systematics could arise for example from biases in the energy reconstruction or other calibration effects, and from detector response nonuniformities.  
Such effects were not significant in EXO-200 and are therefore not expected to significantly impact the sensitivity calculations.

\section{Sensitivity and Discovery Potential Methodology}
\label{sec:sensitivity}

The sensitivity and discovery potential of nEXO are determined by finding the confidence interval of the \0 rate of an ensemble of simulated trial experiments, here called ``toys''. The confidence interval of a toy experiment is determined using a profile likelihood method. In this method, each possible signal rate is treated as a single hypothesis to be tested against the data. Each hypothesis is fit against the data and assigned a likelihood-based test statistic, $\lambda$. The confidence interval is the set of signal rate hypotheses with $\lambda$ values below a threshold determined to produce the desired degree of confidence.
 
To calculate sensitivity, an ensemble of toys is generated assuming zero \0 events, and the upper limit of the 90\% confidence interval is determined for each toy. The sensitivity is the median upper limit of this ensemble. The discovery potential is determined by finding the mean \0 rate  for which 50\% of toys exclude the null hypothesis from the 99.7\% confidence interval.

This method has many similarities to that used to calculate EXO-200's experimental limit, with a significant difference imposed by the improvements in nEXO. The EXO-200 analysis verified that EXO-200's data falls under the 
conditions of validity of Wilks's theorem \cite{Wilks1938}, and therefore took advantage of the consequent statistical simplification: a single threshold value of $\lambda$ defines the 90\% confidence interval, referred to as $\lambda_c$.

nEXO's data will include fewer background events than EXO-200, thus precluding the use of Wilks's theorem which provides an accurate approximation in the limit of large event rates. As a result, nEXO must directly calculate a separate value for the threshold $\lambda_c$ at each hypothesis to include or exclude a given fit result from the confidence interval. The method used here relies on the techniques described in Ref.~\cite{Feldman:1997qc}, which shows how to set confidence intervals correctly according to the frequentist definition while operating in limits where statistical simplifications cannot be relied upon.

There are two phases to calculating the confidence intervals for nEXO simulated data. In the first phase, the distribution of the test statistic $\lambda$ is calculated for  each signal hypothesis $\mu$. This determines $\lambda_c$, the critical value to include a given hypothesis in an experiment's confidence interval. The distributions are calculated for a specific experimental duration (the ``live time'') and background expectation, and can be used to find the limits for any experiment with the same live time and backgrounds. In the second phase, an ensemble of toy experiments is generated and the confidence interval is calculated for each experiment using the previously determined $\lambda_c$. The median upper limit of the confidence intervals of and ensemble generated under the null hypothesis is nEXO's sensitivity.

\subsection{Generating Toy Experiments\label{sec:draw-toys}}
Toy datasets are generated by randomly sampling the probability distributions functions (PDFs), $S_j^\text{SS,MS}$, describing the energy versus distance-to-surface distribution in the detector arising from each background component $j$, with SS and MS PDFs considered separately. When evaluating $\lambda_c(\mu)$ for $\mu>0$, the toy datasets also sample the \0 PDF.

These PDFs are created from the distribution in energy versus distance-to-surface space of simulated backgrounds generated by nEXO's \textsc{Geant4} MC and reconstruction. 
Following the EXO-200 analysis, an energy threshold of 800 keV is used. 

The overall normalization $n$ of each component $j$'s PDF was set using the formula:
\begin{equation}
n^{\text{SS,MS}}_{j} =  M_j\cdot \varepsilon_j^{\text{SS,MS}} \cdot A_j  T
\label{eq:mean_expected_counts}
\end{equation}
where $M$ is either the mass or the surface area of the detector component (whichever is relevant), $\varepsilon$ is the hit efficiency, $A$ is the specific activity for the nuclide, and $T$ is the observation time. Except for $T$, all entries in Eq.~\eqref{eq:mean_expected_counts} are component-specific.

The hit efficiency $\varepsilon^{\text{SS,MS}}$ is the probability that a decay in a specific detector component will produce an event of given multiplicity within the selected region of energy and distance to surface. $\varepsilon^{\text{SS,MS}}$ is obtained from nEXO's \textsc{Geant4} MC, as described in Sec.~\ref{sec:MC-recon}. The number of MC events generated was chosen to keep the statistical uncertainty in the hit efficiency in the inner LXe volume and near the \0 $Q$ value within $\sim20$\% for all significant background contributors. 

Since their true values are unknown, the values for the activities $A_j$ are sampled for each toy experiment from Gaussian distributions truncated at zero and with parameters given by the measured central value and uncertainty from the radioassay measurements in Table~\ref{tab:rad}.  One random draw is performed per material, such that the value of $A_j$ is the same for all background components made of the same material. For radioassays that returned only upper limits, the Gaussian distribution is assumed as centered at zero with the appropriate uncertainty to match the limit. This method introduces a Bayesian element to what is otherwise a frequentist limit-setting methodology by describing the uncertainties in isotope activities with prior probability distributions.  

Equation ~\eqref{eq:mean_expected_counts} and the $S_j^\text{SS,MS}$ are combined to produce a total background spectrum PDF in energy vs distance-to-surface space. This PDF is then randomly sampled to produce a toy dataset, represented by two histograms for the SS and MS events.

Evaluating the $\lambda_c(\mu)$ curve requires $\sim100,000$ toy datasets, at varying values of $\mu$. Once the curve is known, $10,000$ toys with $\mu=0$ are used to estimate the sensitivity. 

\subsection{Fitting Toy Data}

Each toy dataset is fit by minimizing the negative log-likelihood (NLL, $\mathcal{L}$) constructed from the MC-generated PDFs of each component:
\begin{equation}
\mathcal{L}_{\text{nEXO}} = \mathcal{L}_{\text{SS}}+\mathcal{L}_{\text{MS}} - \ln (G_\text{const})
\label{eq:NLL}
\end{equation}
where $\mathcal{L}_{\text{SS(MS)}}$ is the binned NLL built from the toy SS (MS) data compared to the corresponding PDFs $S_j^\text{SS,MS}$ generated by the nEXO MC for each background $j$ (and the \0 signal). The definition of the logarithmic likelihood function closely follows that outlined in Sec.~6.2 of Ref.~\cite{Albert:2013gpz}. The fit parameters are the expected number of counts in each component, $n^{SS+MS}_j$ (total, both SS and MS) and the fraction of SS events in that component $f^\text{SS}_j$. These parameters are fit, rather than being fixed, to accommodate the uncertainty the final nEXO experiment will have about background intensities and SS/MS discrimination.

$G_\text{const}$ is a multivariate Gaussian function constraining some fit parameters. 
The SS fractions $f_\text{SS}^j$ for all components are constrained to be within 5.9\% of the mean obtained by the MC simulations, as per Ref.~\cite{Albert:2013gpz}. The background rate from \ce{^222Rn} in the LXe outside of the TPC region is constrained to be within 10\% of the expectation rate, as this will be well known during the experiment by studying tagged \ce{^222Rn} in the TPC volume. The normalization terms $n^{SS+MS}_j$ are unconstrained. The covariance matrix is diagonal in this study, as systematics causing correlation between data bins have not yet been taken into account.

The NLL fit is implemented using \textsc{RooFit} \cite{Verkerke:2003ir} and \textsc{Minuit} \cite{Lazzaro:2010zza}. 

\subsection{Calculating the Test Statistic Distribution}

The NLL ratio test statistic is calculated as
\begin{equation}
\lambda(\mu) = 2 \left( \mathcal{L}_\mu - \mathcal{L}_{\mu_\text{best}} \right)
\end{equation}
where $\mathcal{L}_\mu$ is the log-likelihood fixing the signal expectation to $\mu$ and $\mathcal{L}_{\mu_\text{best}}$ is the log-likelihood letting the signal parameter assume its best-fit value $\mu_\text{best}$. 

Instead of relying on Wilks's theorem, the NLL ratio threshold $\lambda_c(\mu)$ is explicitly computed at all values of $\mu$ covered in this study.

Following the approach suggested in Ref.~\cite{Feldman:1997qc}, the $\lambda_c(\mu)$ curve is obtained in a frequentist manner via MC generation of the distribution of the test statistic under each hypothesis. 
Over a range of hypotheses $\mu$, an ensemble of toy experiments is generated with a number of signal counts randomly drawn from the expectation $\mu$. The value of $\lambda(\mu)$ for each of these experiments is computed and the 90th percentile of the resulting distribution of $\lambda(\mu)$ defines the critical value $\lambda_c(\mu)$ for the 90\% confidence interval used for sensitivity. The 99.7th percentile similarly defines $\lambda_c(\mu)$ for the 99.7\% confidence interval used for the discovery potential.

Calculating $\lambda_c(\mu)$ requires generating and fitting an ensemble of many toys under a range of values for $\mu$. 
To reduce computing time, the calculation is performed at several discrete points which are then fit with a third-order spline. 
This produces a smooth curve that interpolates between the calculated points and reduces the impact of the statistical uncertainty of the calculation of a quantile on a finite distribution.

The $\lambda_c(\mu)$ curve is obtained under a specified live time and expected distribution of backgrounds. Changing either of those assumptions requires the calculation of a new curve.

\subsection{Calculating Sensitivity and Discovery Potential}

\begin{figure}[tbp]
\centering
\includegraphics[width=0.99\columnwidth]{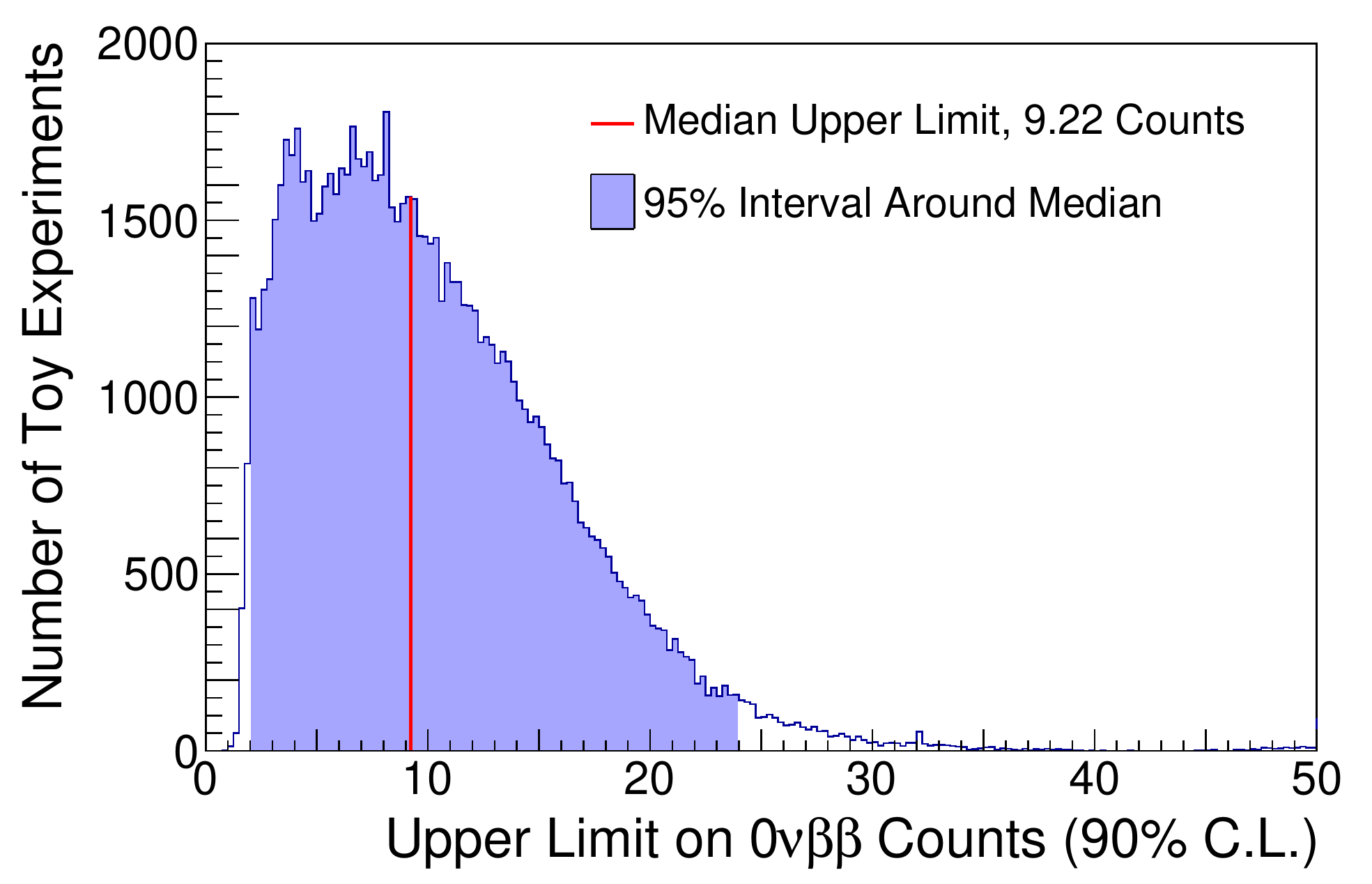}
\caption{Distribution of $\mu_{90}$, the upper limit on the signal counts under the null hypothesis, obtained for several background realizations (toy experiments) at 10-yr livetime.}
\label{fig:ul-sens-dist-10y}
\end{figure}

For any single toy data set, a given value of $\mu$ is included in the set of hypotheses that make up the confidence interval $C$ if $\lambda(\mu)<\lambda_c(\mu)$.
A bisection algorithm is used to minimize the number of $\lambda(\mu)$ points that must be calculated for each toy experiment to determine $\mu_{90}$, the crossing point (or the greater of two crossing points, if two exist) between $\lambda_c(\mu)$ and $\lambda(\mu)$. This approach was validated by comparing the results obtained in the high-statistics regime, where Wilks's theorem holds, against those from \textsc{Minuit}.

nEXO's sensitivity at a given background and live time is extracted as the median of the distribution of the upper limit $\mu_{90}$ from an ensemble of toy experiments generated with the null hypothesis $\mu=0$ and under those livetime and background assumptions. An example of the distribution of $\mu_{90}$ is shown in Fig.~\ref{fig:ul-sens-dist-10y}. The \0 half-life sensitivity is then inferred from the median $\mu_{90}$, the number of \ce{^136Xe} nuclei and the experiment's live time.

In addition to the sensitivity, the median discovery potential at 3-sigma is also calculated. An experiment is a discovery if the 99.7\% confidence interval, calculated as described above, does not include the null hypothesis $\mu=0$. The median discovery potential is the \0 half-life that produces discovery experiments 50\% of the time. Determining the discovery potential entails a search over \0 rates to find which rate produces 50\% discoveries.

\section{Sensitivity and Discovery Potential to \0}
\label{sec:results}

\subsection{Simulated Spectra and Background Budget}

Three parameters---energy, distance to nearest surface, and event type (SS or MS)---are used to characterize signal and background events. 

\begin{figure*}[tbp]
      \centering   
     \includegraphics[width=0.7\textwidth]{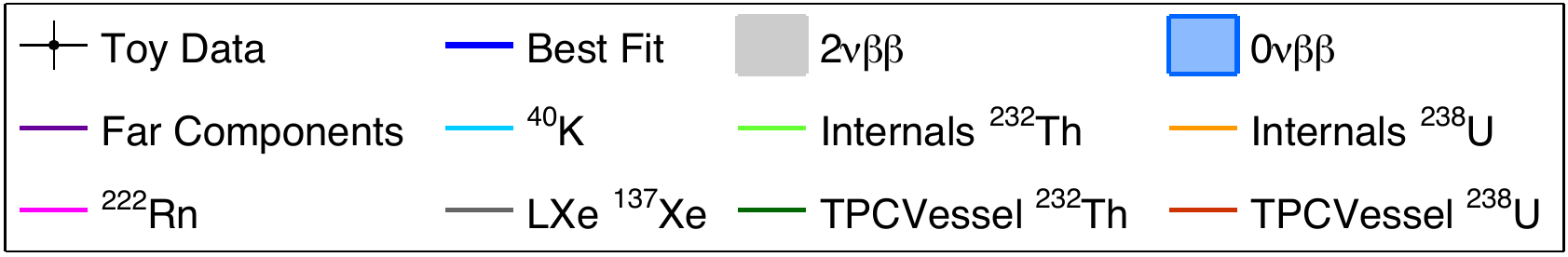} 
\includegraphics[width=0.49\textwidth]{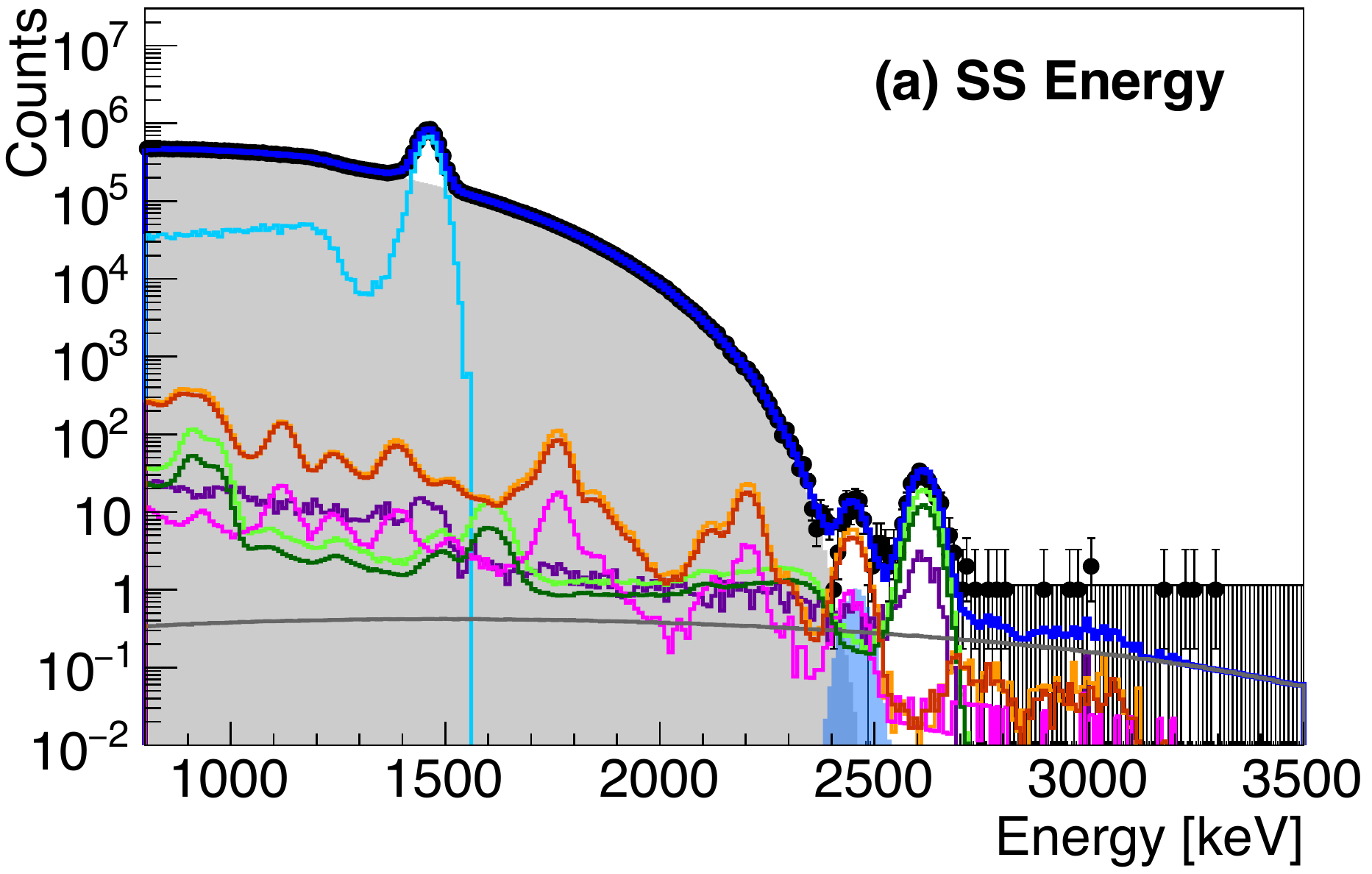}
\includegraphics[width=0.49\textwidth]{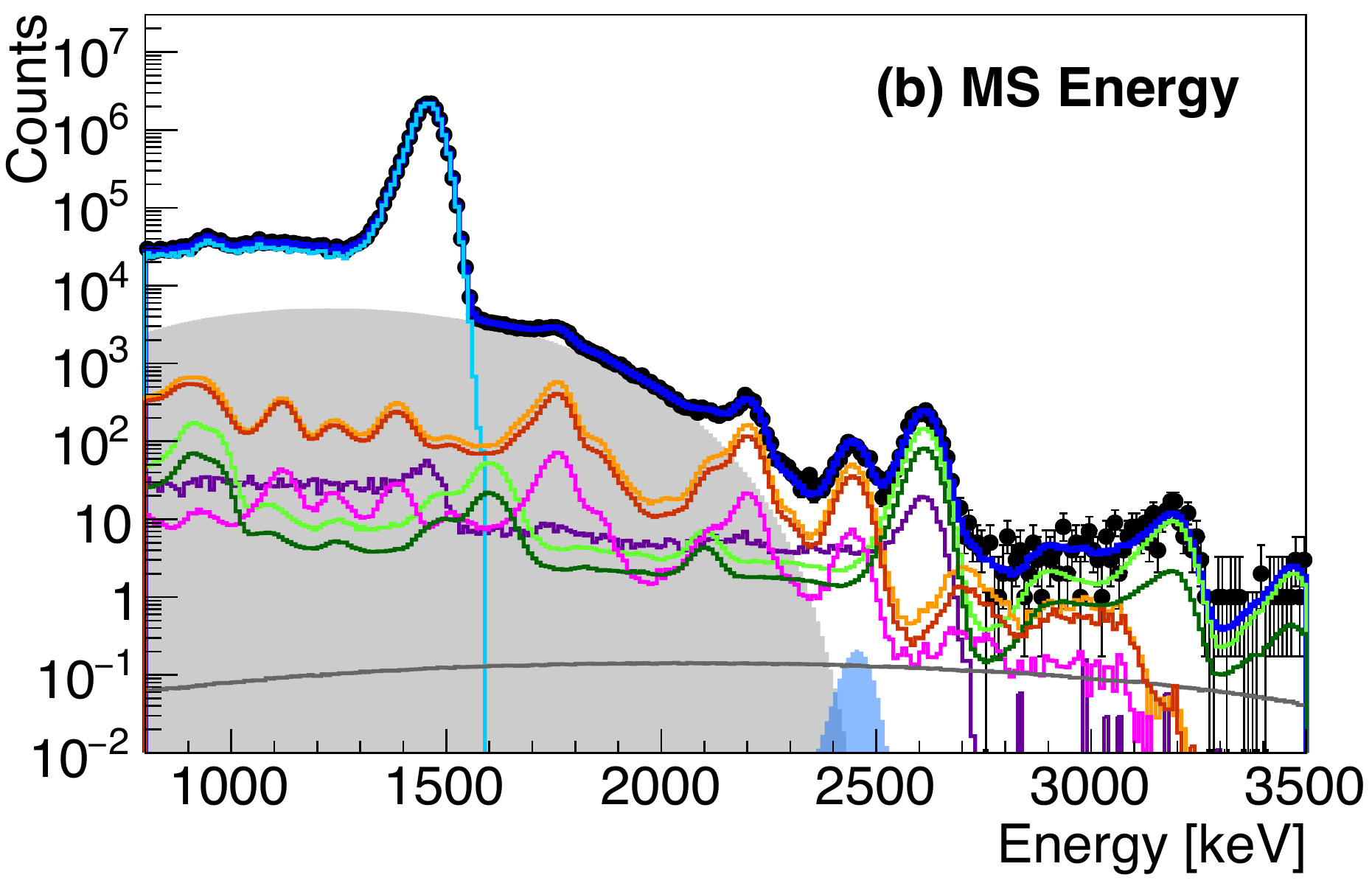}
\includegraphics[width=0.47\textwidth]{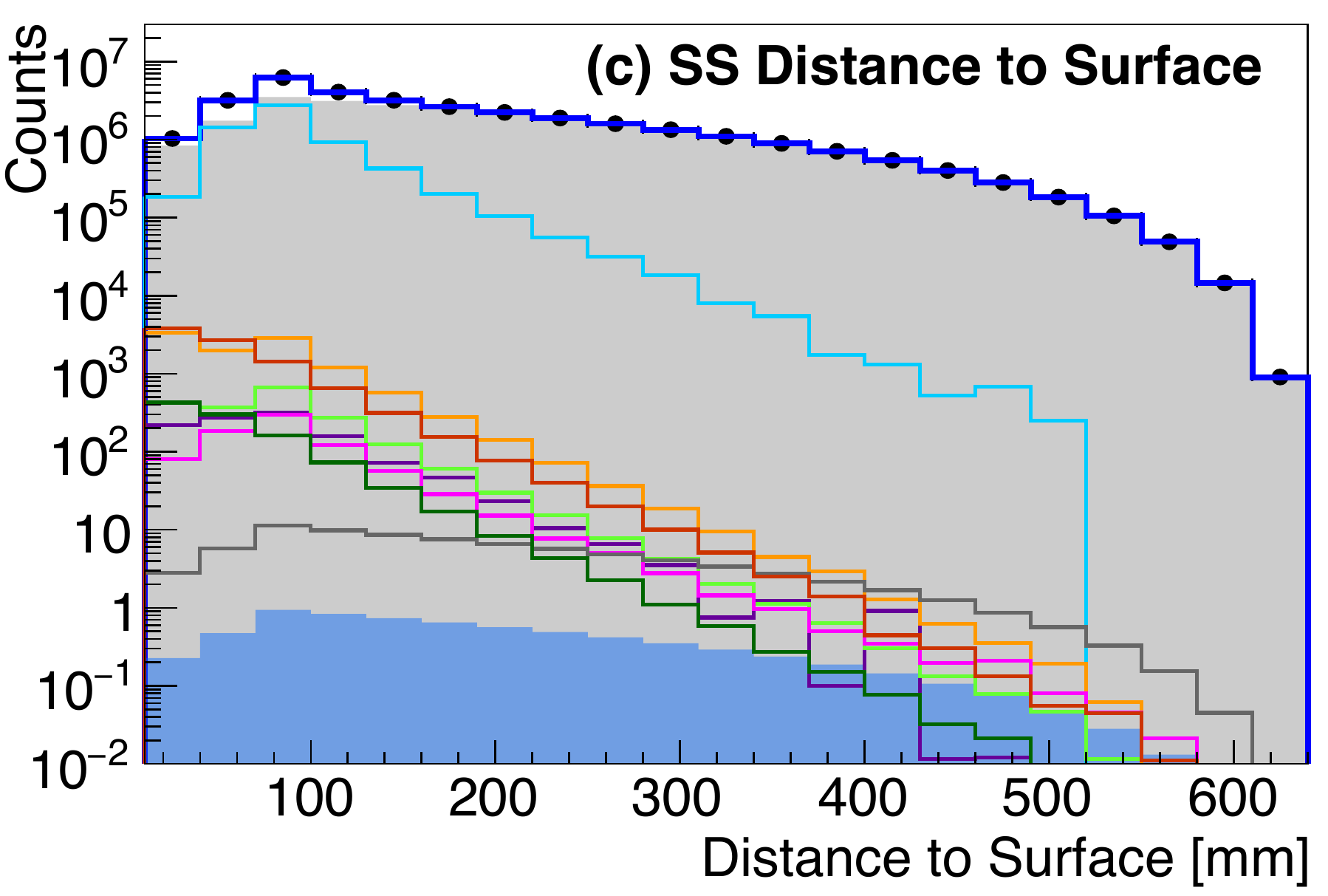}
\hspace{0.2cm}
\includegraphics[width=0.47\textwidth]{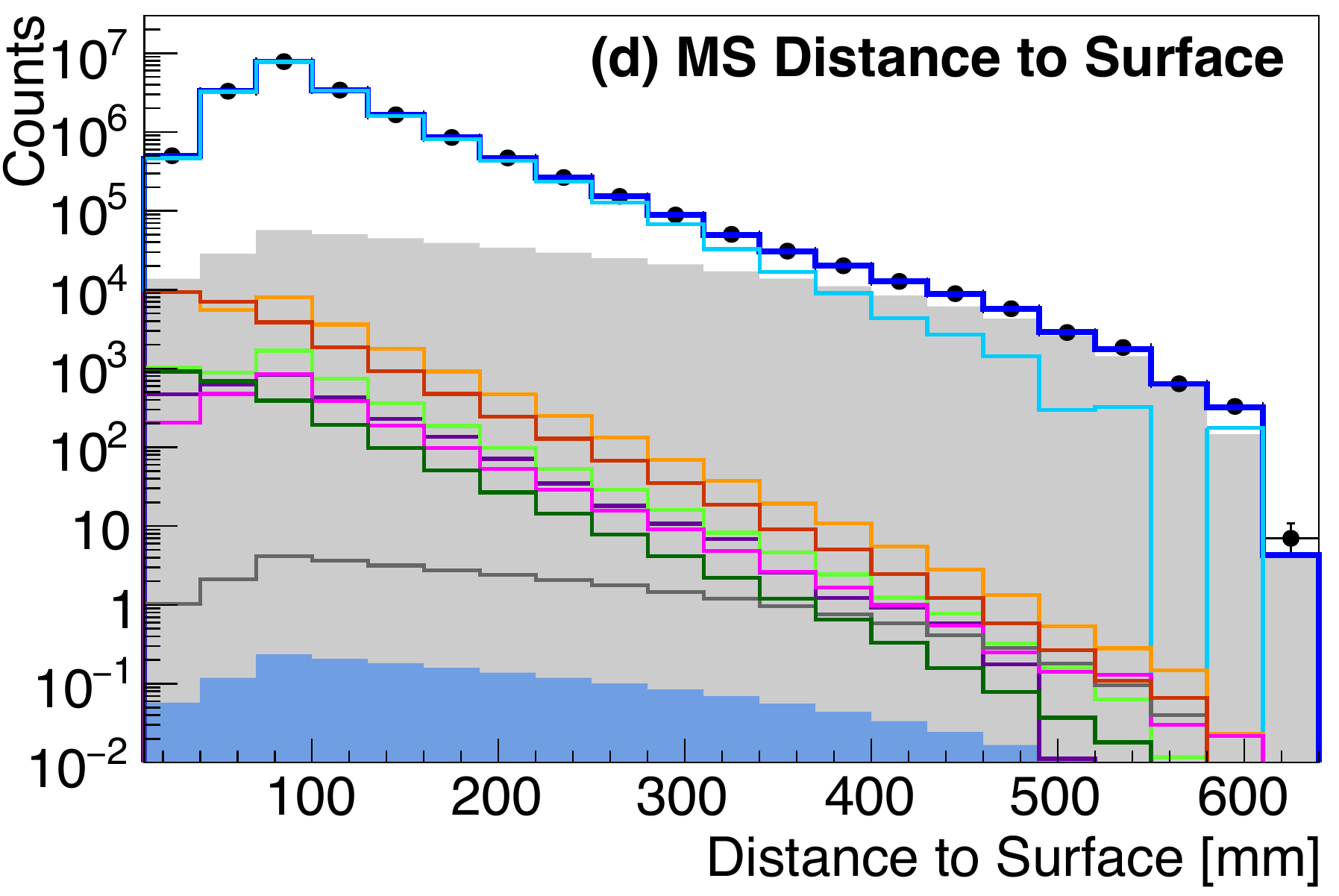}
      \caption{Result of the NLL fit on a representative nEXO toy dataset generated assuming  a \0 signal corresponding to a half-life of $5.7\times 10^{27}$yr and 10 yr of detector live time. The top plots are the energy distribution histograms while the bottom plots are the histograms of distances to surface; left (right) spectra are for SS (MS) events.
}
      \label{fig:BestFitSpectra}
    \end{figure*}

As an example of this multidimensional analysis, the SS and MS spectra from one of the simulated toy experiments are given in  Fig.~\ref{fig:BestFitSpectra}. Best fit results for all radionuclides (arranged by component groups) are shown.  Overall, the energy spectra are dominated by \2 events, while the tail of the \ce{^{214}Bi} photoelectric peak is the largest background contributor near the \0 $Q$ value.  
As a function of distance to surface, the distribution of external radioactivity drops rapidly and is markedly different than the distribution of \0, \2, and \ce{^137Xe} events which are uniform throughout the LXe volume. This behavior adds resolving power other than energy to the analysis, improving the ability to distinguish a \0 signal from background. 

\begin{figure*}[tbp]
\centering
\includegraphics[width=0.7\textwidth]{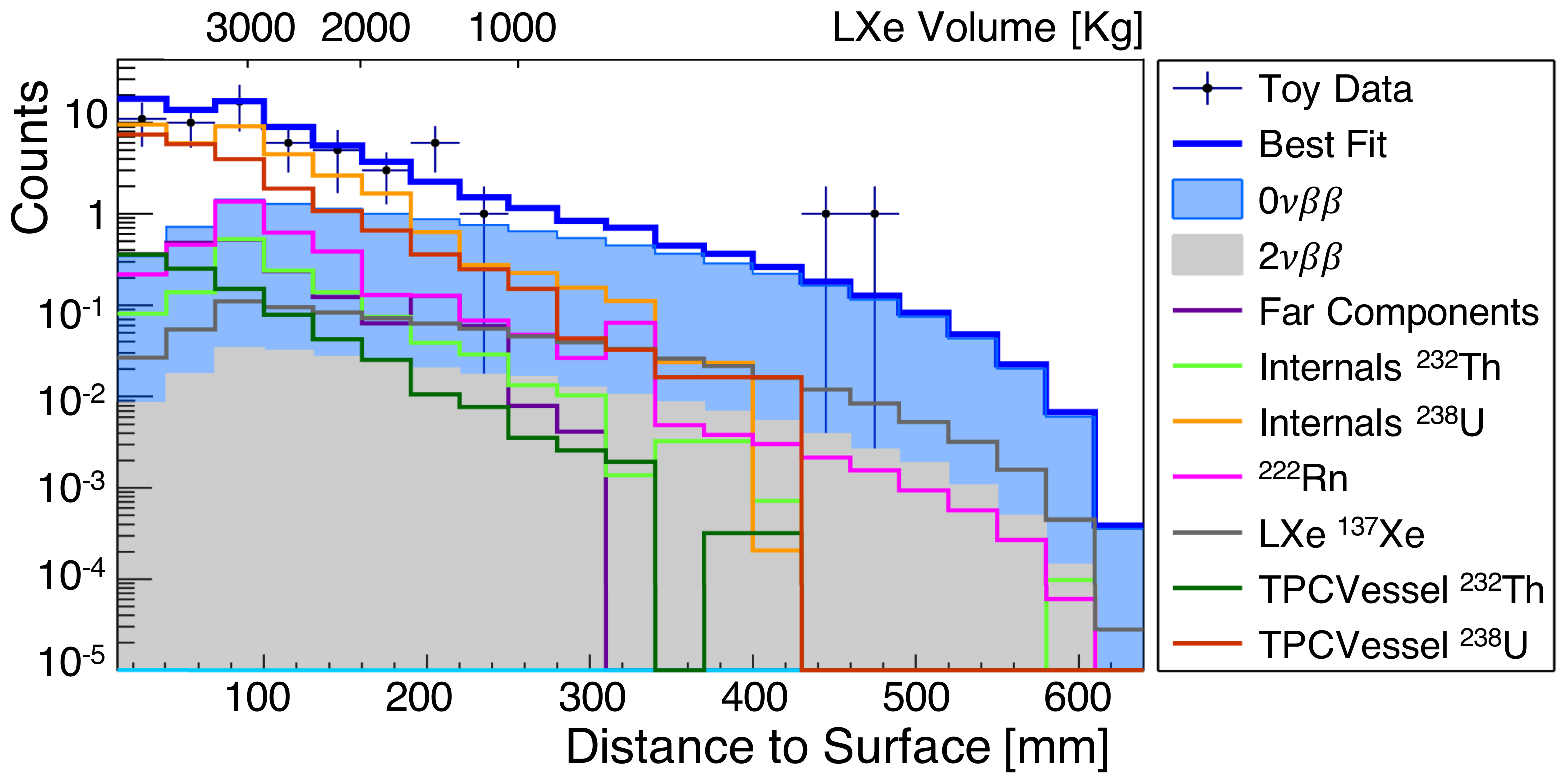}
\caption{Distance to surface distribution for the fit results from a representative toy MC dataset with 10 yr livetime. Only SS events with energy within \Q$\pm$FWHM/2 are included.  The \0 signal corresponds to a half-life of $5.7\times 10^{27}$yr.}
\label{fig:Fit_SS_Standoff_ROI} 
\end{figure*}

To highlight the backgrounds of greatest concern, the  distance to surface distribution is shown in Fig.~\ref{fig:Fit_SS_Standoff_ROI} with a cut selecting the $\pm$FWHM/2 energy region around \Q.

\begin{figure*}[tbp]
      \centering
\includegraphics[width=0.49\textwidth]{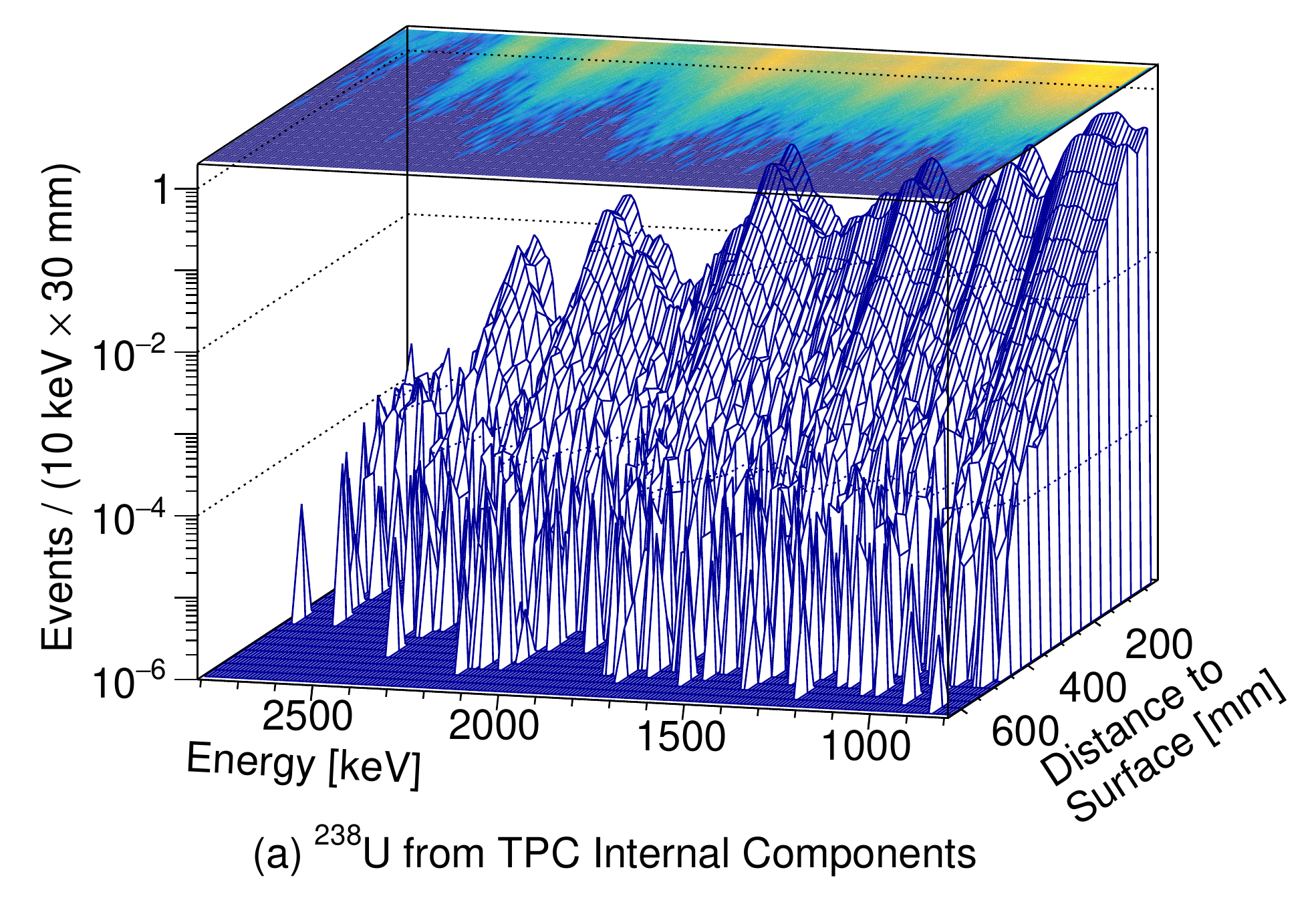}
\includegraphics[width=0.49\textwidth]{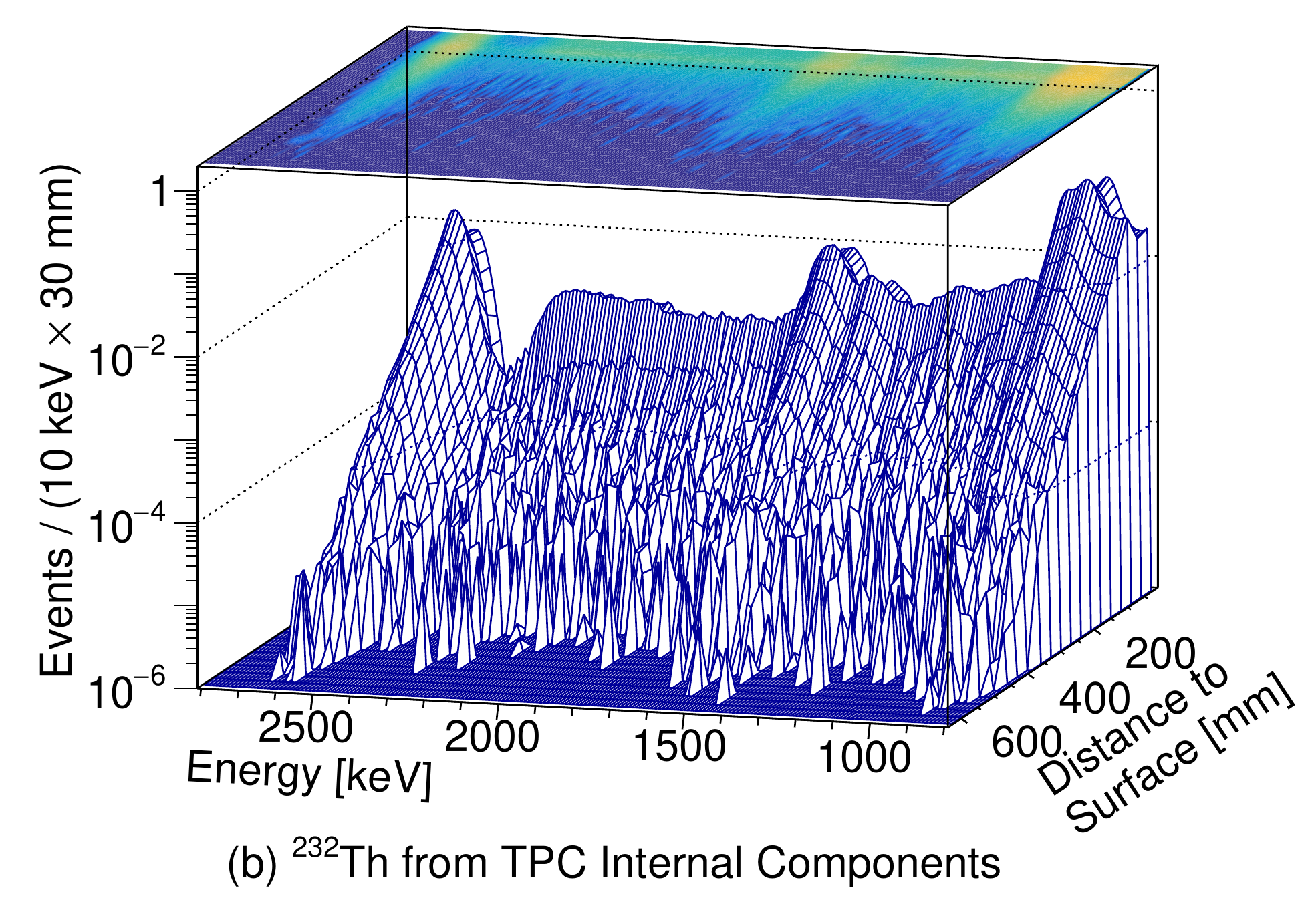}
\caption{Distribution of SS counts for a representative toy MC experiment as a function of energy and distance to surface for the \ce{^238U} (a) and \ce{^232Th} (b) background arising from all TPC internal components. }
\label{fig:3Dspectra}
\end{figure*}

In the central region of the detector, external gamma backgrounds are reduced by several orders of magnitude (Fig.~\ref{fig:3Dspectra}). Given the size of nEXO and the absence of any material other than LXe within the TPC volume, 2.5 MeV gamma-rays have to traverse more than seven attenuation lengths, and likely scatter multiple times, before reaching the center of the LXe volume.  

\begin{figure*}[tbp]
\centering
\includegraphics[width=0.8\textwidth]{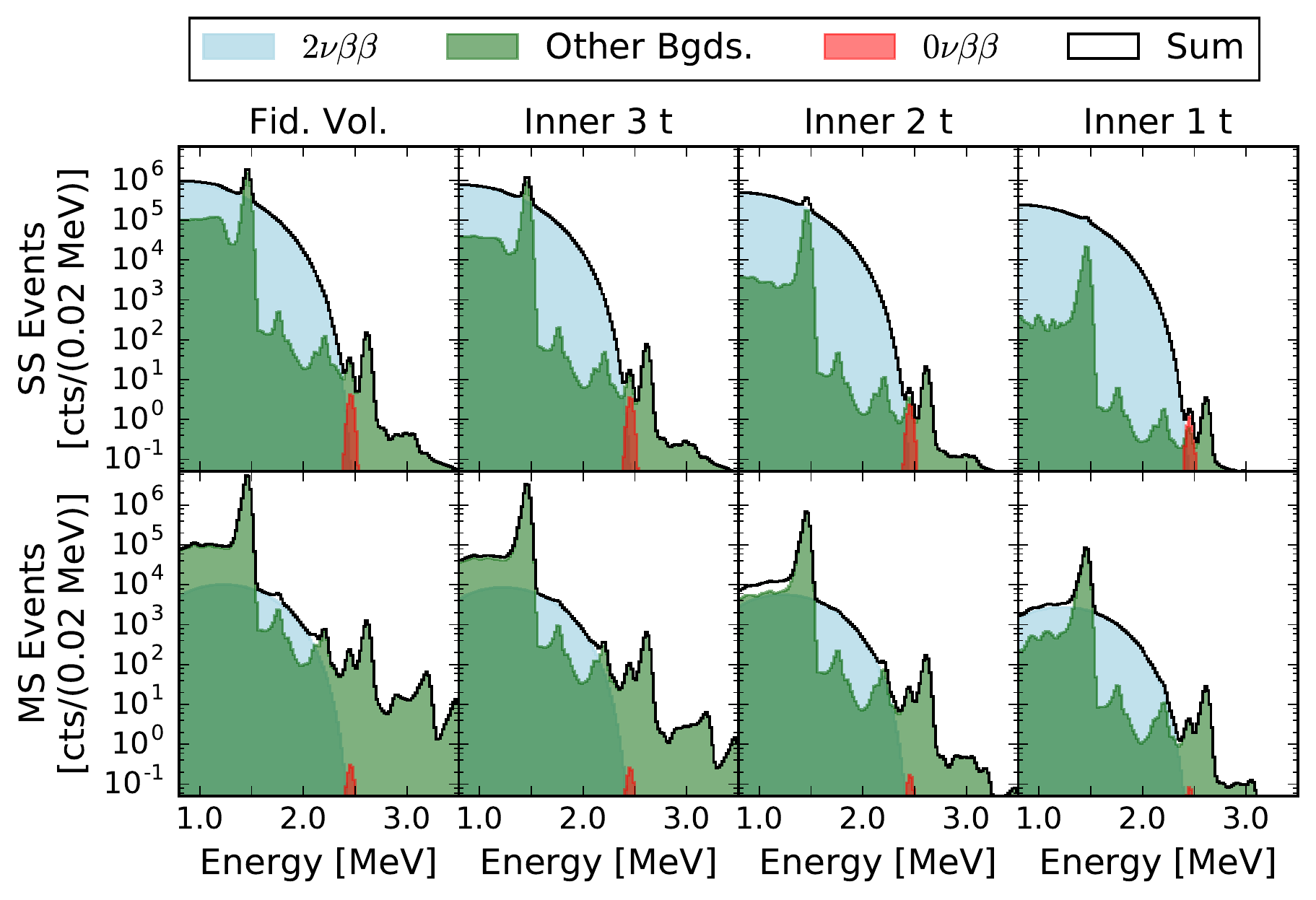}
\caption{Energy spectra for SS and MS events as a function of the LXe mass. Spectra are evaluated for a detector livetime of 10 yr. The \0 signal corresponds to a half-life of $5.7\times 10^{27}$ yr.}
\label{fig:SS-MS-inner_volumes}
\end{figure*}

The diagnostic power of nEXO's multiparameter approach can be further appreciated by looking at Fig.~\ref{fig:SS-MS-inner_volumes}. While the energy resolution alone marginally resolves a \0-peak from $\gamma$-peaks caused by external radioactivity, the distance to surface variable provides the additional resolving power in combination with the event-type variable (SS/MS). The large body of xenon is not simply used as a passive shield but actively measures external backgrounds and internal double-$\beta$ decays simultaneously. The outer volumes effectively quantify external backgrounds, while the inner volumes determine the $\beta\beta$-signal. This combination of variables adds confidence in case of a discovery. 

\begin{figure}[tbp]
	\centering
	\includegraphics[width=0.95\columnwidth]{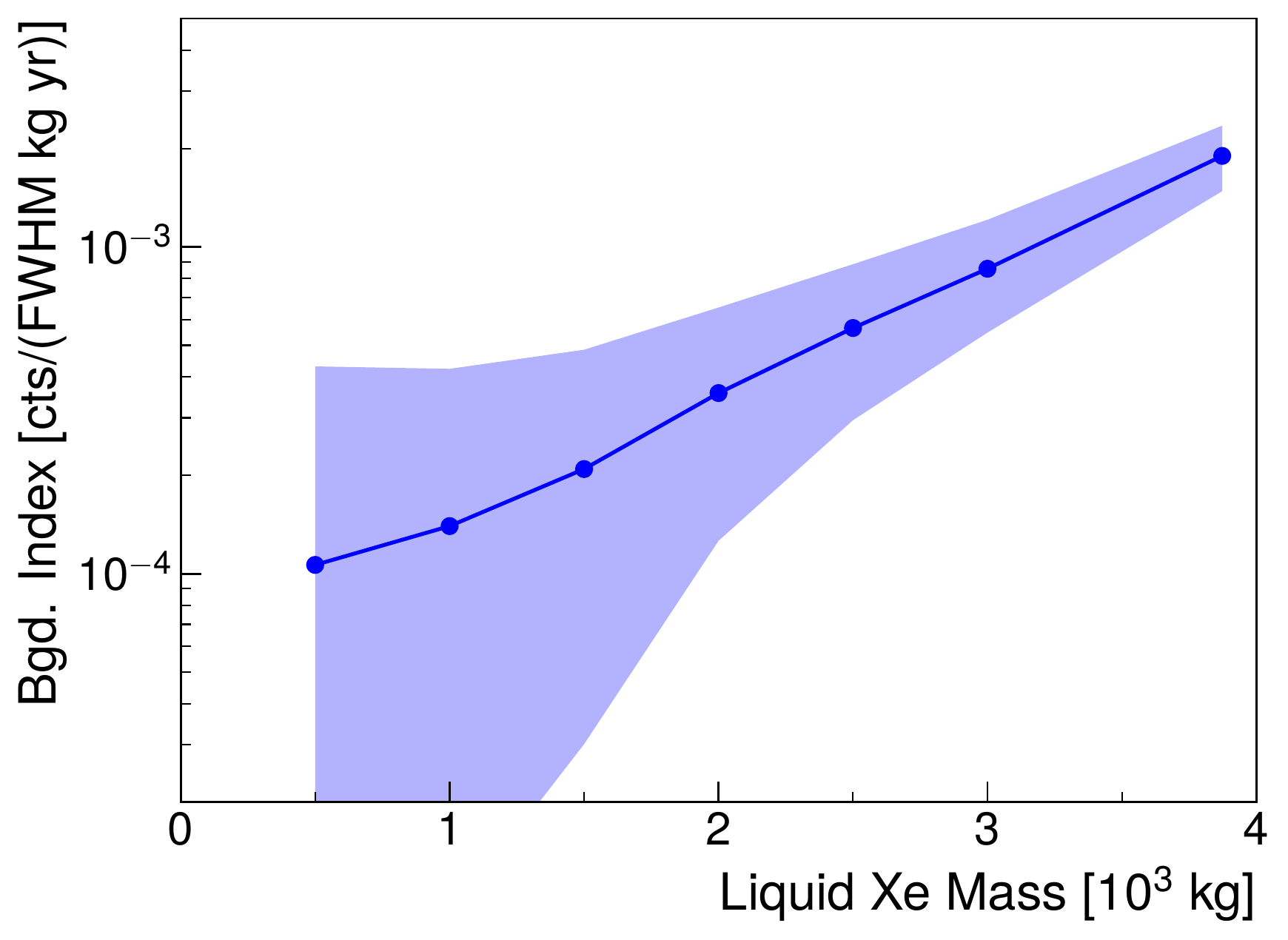}
	\caption{Median background as a function of the LXe fiducial mass derived from $10^4$ toy-MC simulations with detector live time of 10 years. The band corresponds to the 95\% confidence belt of the distribution of background counts at each fiducial mass value.}
	\label{fig:bkd_index_vs_fiducial_mass}
\end{figure}

The count rate as a function of the fiducial mass is shown in Fig.~\ref{fig:bkd_index_vs_fiducial_mass} for events in an energy window $\pm$FWHM/2 around \Q. Shown is the median and 95\% band of the distribution resulting from the random draw of activities described in Sec.~\ref{sec:draw-toys}. 
Clearly, a large homogeneous detector like nEXO cannot be characterized by a single background index value. Instead, its background rate is a position-dependent function. 
While for specific and circumscribed purposes it may be convenient to think in terms of a single background rate in a region of energy, one should always be aware that this  point of view is not generally appropriate here.
However, as a reference, nEXO is predicted to achieve a median background rate of $3.6\times10^{-4}$ cts/(FWHM$\cdot$kg$\cdot$yr) in the inner 2000 kg of LXe. This choice of mass value will become clear in the next section.

\begin{figure*}[tbp]
\centering
\includegraphics[height=6.6cm]{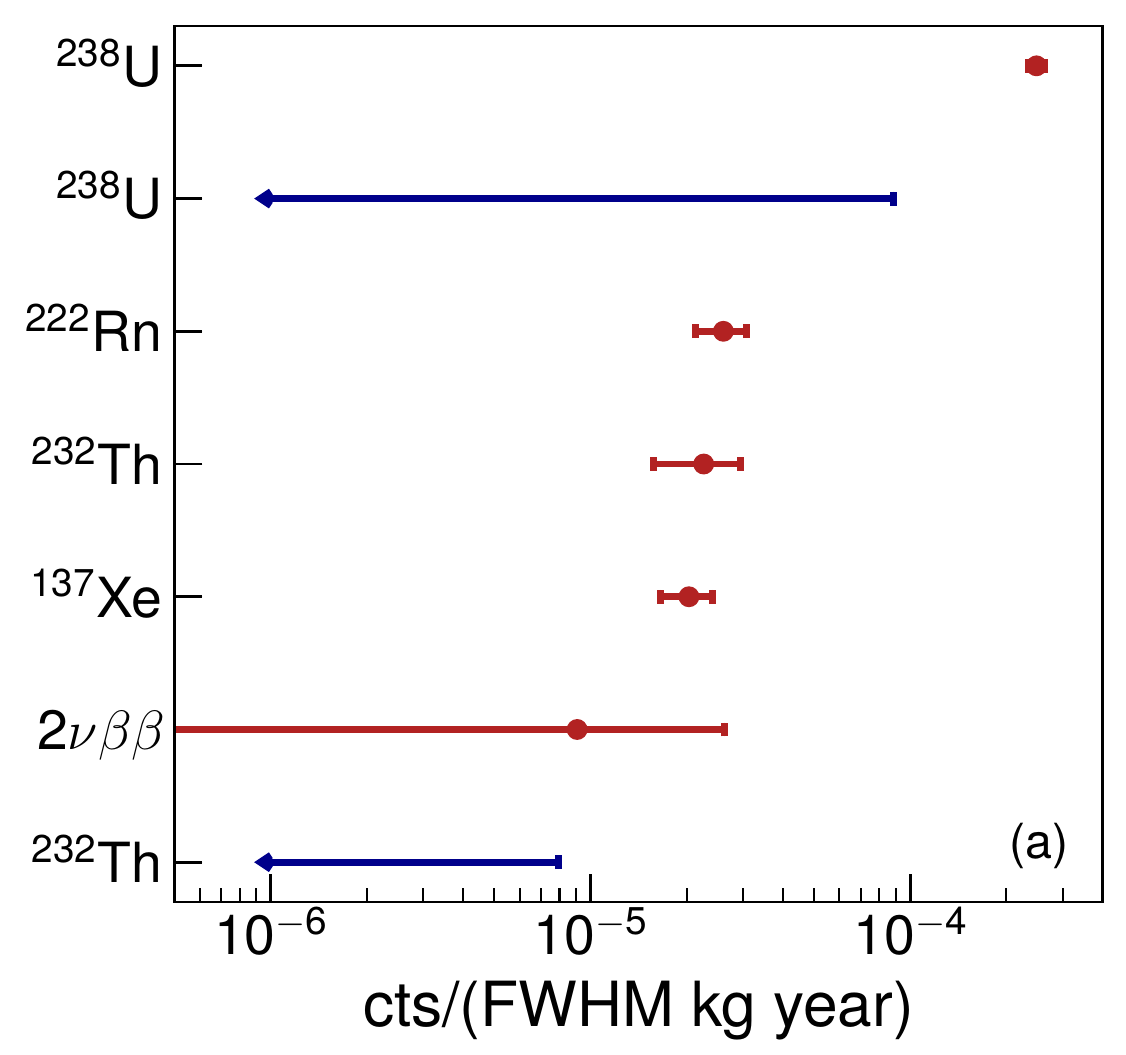}
\includegraphics[height=6.6cm]{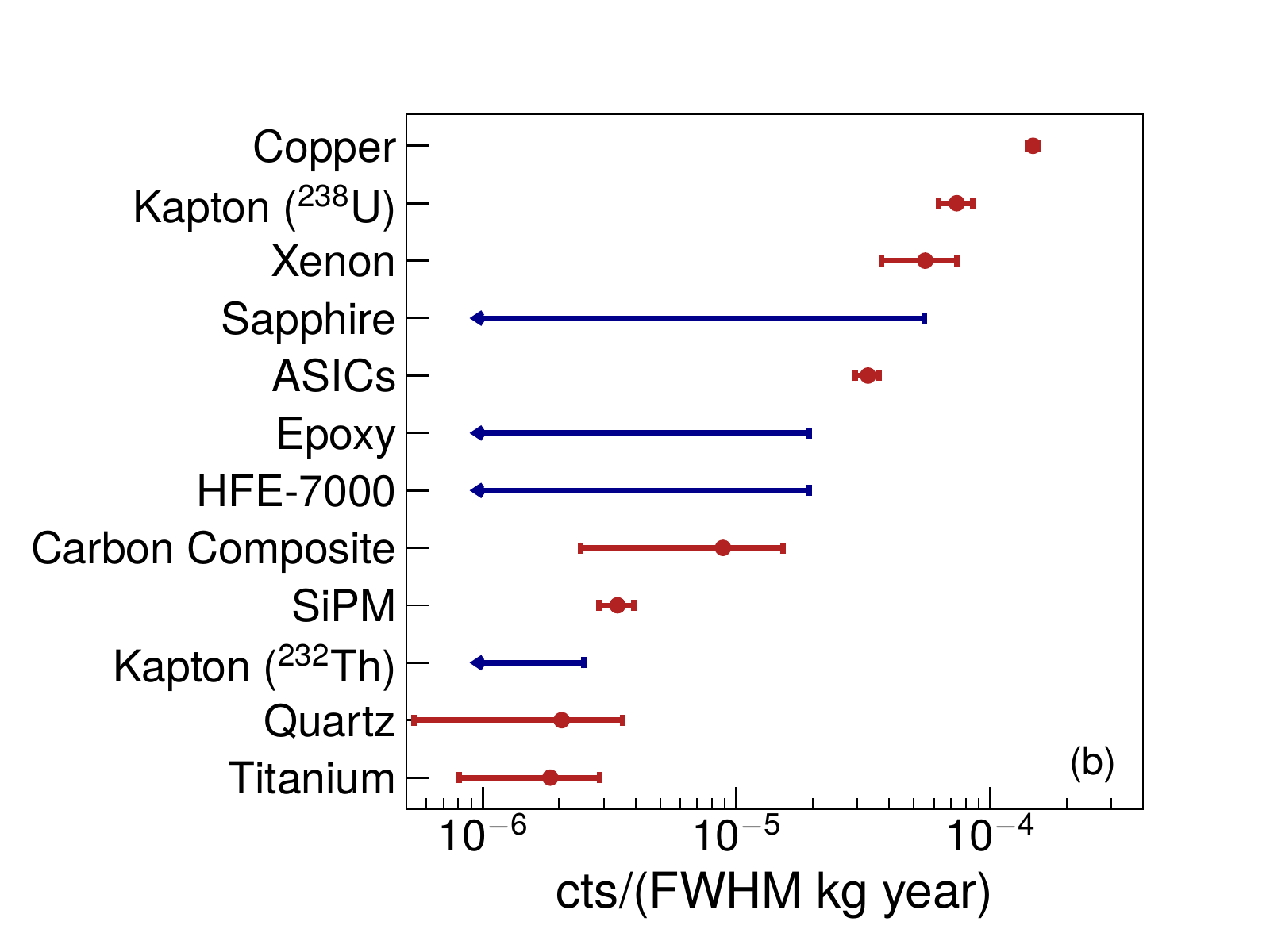}\\
\vspace{0.5cm}
\includegraphics[width=0.7\textwidth]{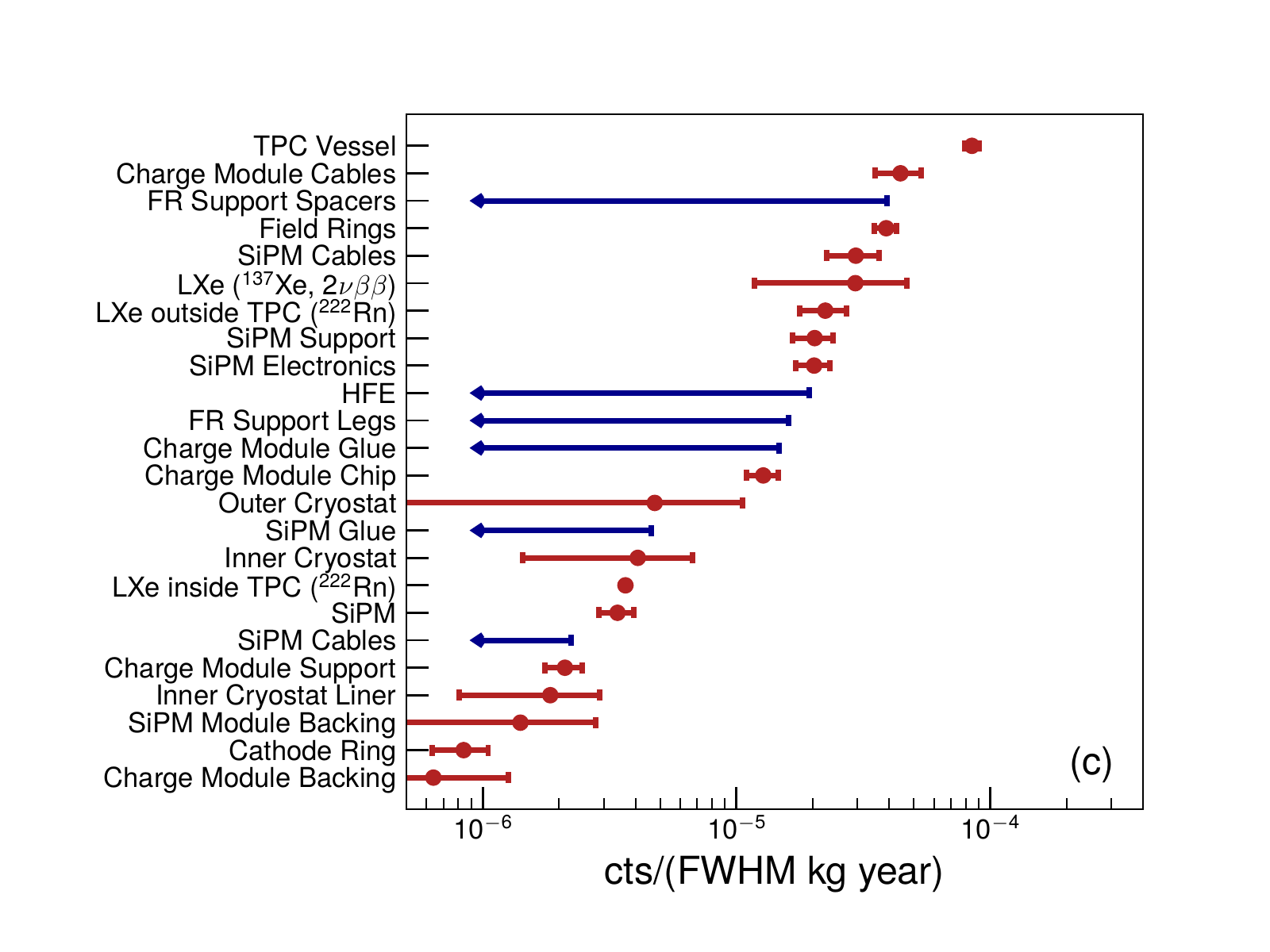}
\caption{Histograms of the SS background contributions by nuclide (a),   material (b), and detector component (c). for nEXO with energy within \Q$\pm$FWHM/2 and in the inner 2000 kg. The arrows indicate 90\% C.L. upper limits while the circles indicate measured values with 1$\sigma$ uncertainties. 
Systematic uncertainties and contributions smaller than $5\times10^{-7}$ cts/(FWHM$\cdot$kg$\cdot$yr) are not shown. }
	\label{fig:nEXO_background_budgets}
\end{figure*}

Figure \ref{fig:nEXO_background_budgets} visualizes the contribution of different background components and their uncertainties. This figure shows the measured activities in Table \ref{tab:rad} multiplied by the SS hit efficiency for events with energy within \Q$\pm$FWHM/2 and within the inner 2000 kg of LXe. Contributions to the background budget are grouped by material, nuclide, and component. Contributions for which a measured radioassay exists are considered separately from those for which only an upper limit is available.

$\gamma$-ray interactions from the \ce{^{238}U} decay chain constitute more than 70\% of nEXO's background in the inner 2000 kg. A fraction of this component arises from materials for which only upper limits are currently available. Hence nEXO's expected background may fall as these radioassays are replaced by more precise measurements or higher purity materials are selected for use in the nEXO design.

Improvements in the data analysis are also expected to reduce the background arising from \ce{^137Xe}. 
This is important because, as shown in Fig.~\ref{fig:Fit_SS_Standoff_ROI}, this background is uniformly distributed in the detector volume.
At a sufficiently deep location, a straightforward active muon veto could efficiently reduce the \ce{^137Xe} with acceptable loss in livetime. As a conservative measure, no muon-based vetoing has been assumed in the analysis presented here.

The breakdown  by component in Fig.~\ref{fig:nEXO_background_budgets} shows that TPC elements dominate the background due to their vicinity to the central LXe region, while massive but distant components such as the cryostat vessels are subdominant. By material, the largest contribution arises from radio-impurities in the copper, primarily in the TPC vessel which is the largest-mass component near the LXe. Cables and field rings (and their associated support equipment) are the next largest components.  Overall, background counts are rather evenly distributed across various TPC internal components. This is indicative of a well-balanced experimental design.

\begin{figure*}[tbp]
\centering
\includegraphics[width=0.75\textwidth, angle=0]{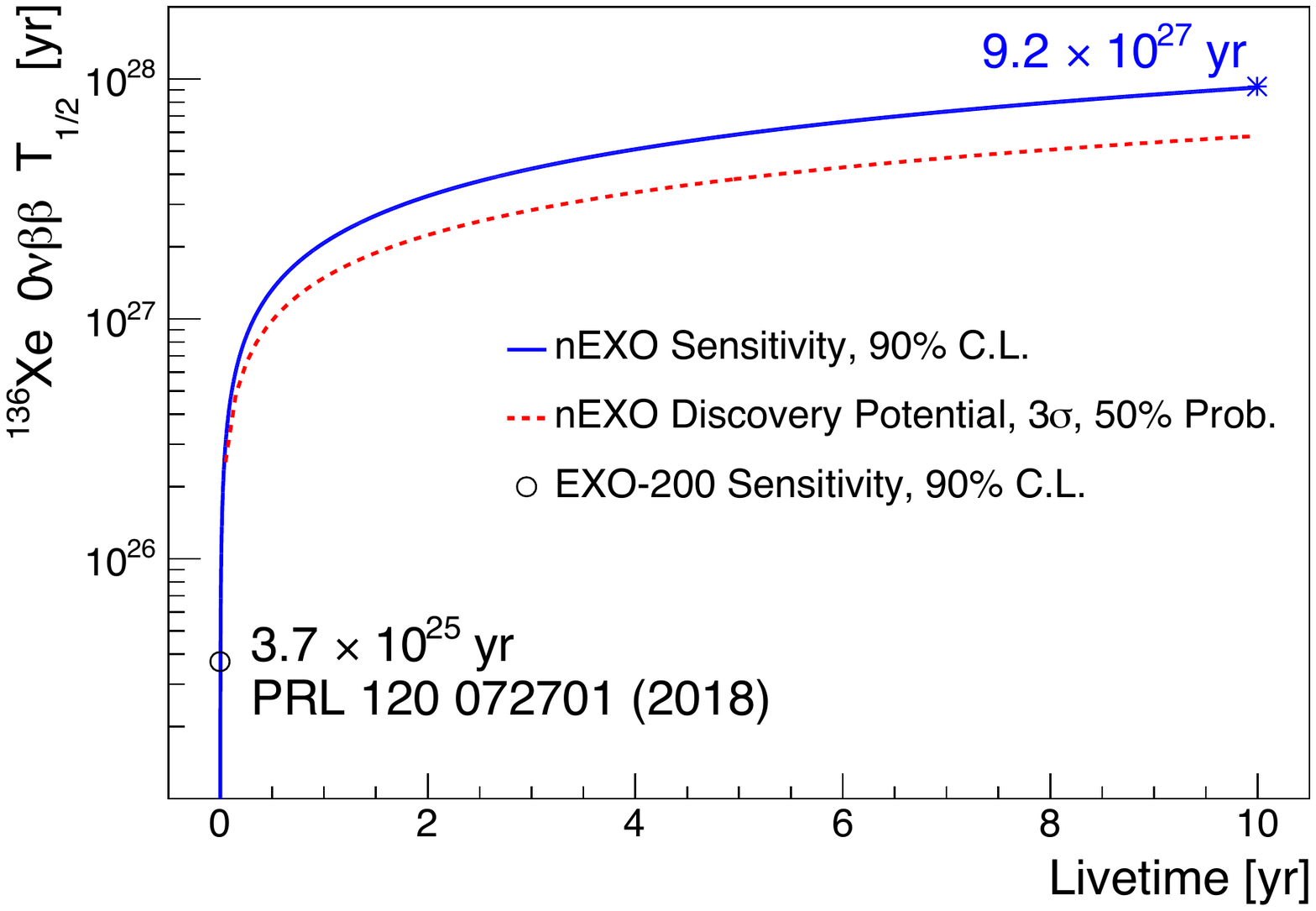}
\caption{\label{fig:nom_sen} nEXO median sensitivity at 90\% C.L. and 3$\sigma$ discovery potential as a function of the experiment livetime. }
\end{figure*}

\subsection{Sensitivity Results}
\label{sec:sub_sens_results}

nEXO's size and extremely low background levels, coupled with the fit analysis, which exploits the multiparameter event signature provided by the TPC technique, result in a dramatic improvement in sensitivity compared to EXO-200 achievements.
nEXO's median sensitivity on the \0 half-life for \ce{^{136}Xe} at 90\% C.L. 
is shown in Fig.~\ref{fig:nom_sen} as a function of the experiment's live time. After 10 years of data collection, the median 90\% C.L. sensitivity reaches $9.2\times10^{27}$ yr. A $3\sigma$ discovery potential of $5.7\times10^{27}$ yr is predicted for the same live time.
  
\begin{figure}[tbp]
\centering
\includegraphics[width=0.95\columnwidth]{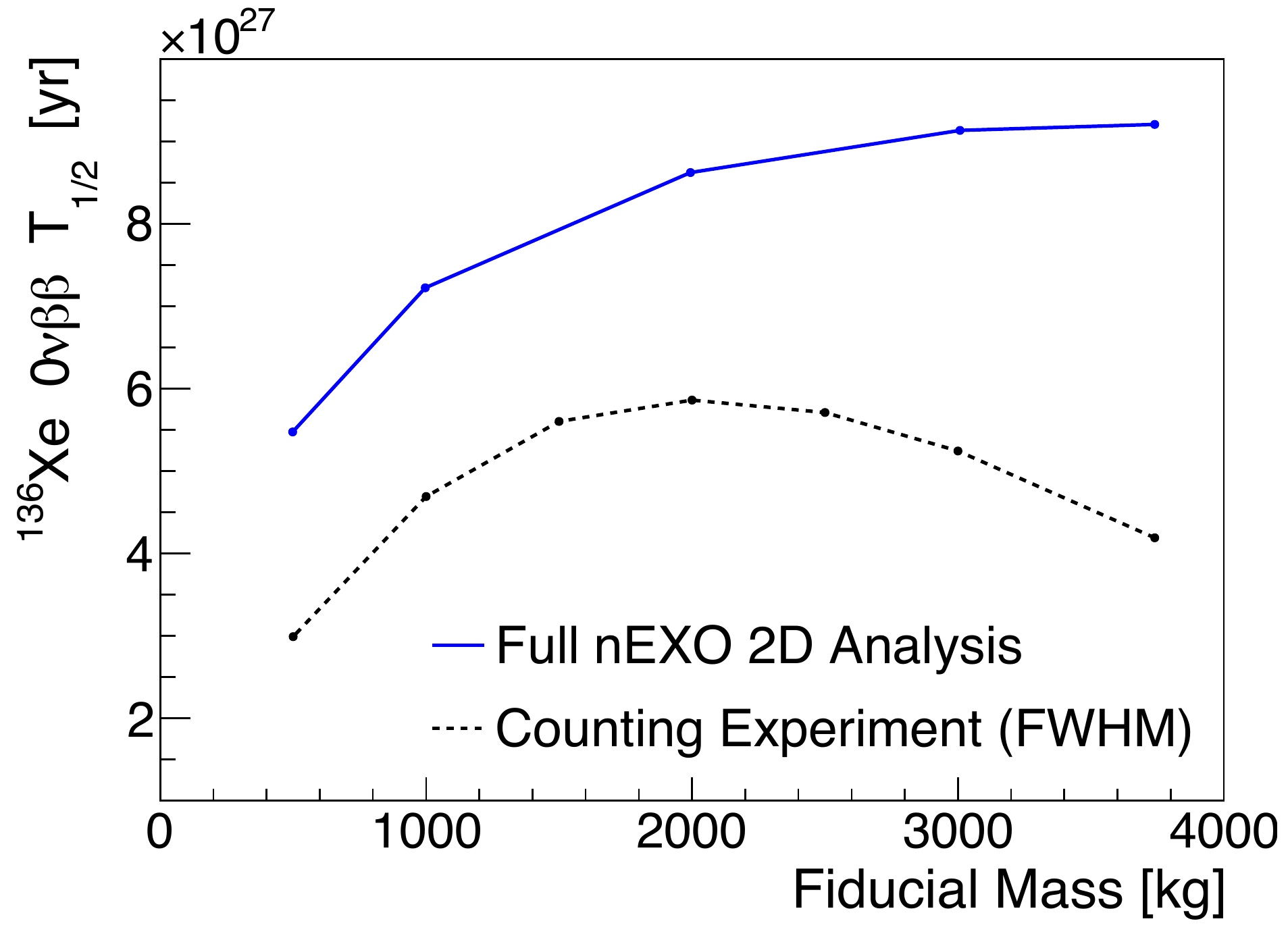}
\caption{\label{fig:sen_vs_fv} nEXO exclusion sensitivity at 90\% C.L. as a function of fiducial LXe volume. The blue points (upper curve) are obtained from the full 2D fit of energy vs distance to surface, while the black points (lower curve) are the result of a pure counting experiment of events with energy in \Q$\pm$FWHM/2. Both analyses are performed using the method of Ref.~\cite{Feldman:1997qc}. }
\end{figure}

The two-dimensional fit of energy and distance to surface allows nEXO to maximize its sensitivity by employing the largest possible fiducial volume, in contrast to a counting analysis which reaches maximum sensitivity only with a substantial fiducial volume cut. This is shown in Fig.~\ref{fig:sen_vs_fv}. Indeed, the full two-dimensional analysis shows an improvement of $\sim$50\% over a counting-style experiment.
Fig.~\ref{fig:sen_vs_fv} also motivates the earlier choice of presenting the background rate for the innermost 2000 kg of LXe where $\sim$90\% of the full sensitivity is achieved and a counting-style rate analysis reaches its maximal sensitivity.

\subsection{Sensitivity Variation Studies}

The results presented are based on a robust estimation of the backgrounds and realistic detector performance, extrapolated from EXO-200 and supported by nEXO-specific modeling results. We note that these results do not involve any extrapolation of materials radiopurity beyond what has been already measured. As the R\&D continues, it is possible that better performance might be achieved e.g. through improved material selection and engineering, improved analysis techniques, or hardware breakthroughs. 

Traditionally, the analysis of energy spectra alone has been the workhorse of \0 searches, thus favoring calorimetric experiments with very high energy resolution. Over time, all \0 searches have started to introduce multiple parameters to measure and reject backgrounds (see, e.g., Refs.~\cite{Petry:1993qp,Hellmig:2000xp,Ferrario:2015kta}). By including  distance to surface  and event type (SS/MS) parameters, EXO-200 was able to provide outstanding physics results in spite of a somewhat limited energy resolution. nEXO's larger mass enhances the utility of these additional variables in the multiparameter analysis, further lessening the reliance on energy resolution. The homogeneous nature of the detector permits to take full advantage of this multiparameter approach.

\begin{figure}[tbp]
\centering
\includegraphics[width=0.95\columnwidth]{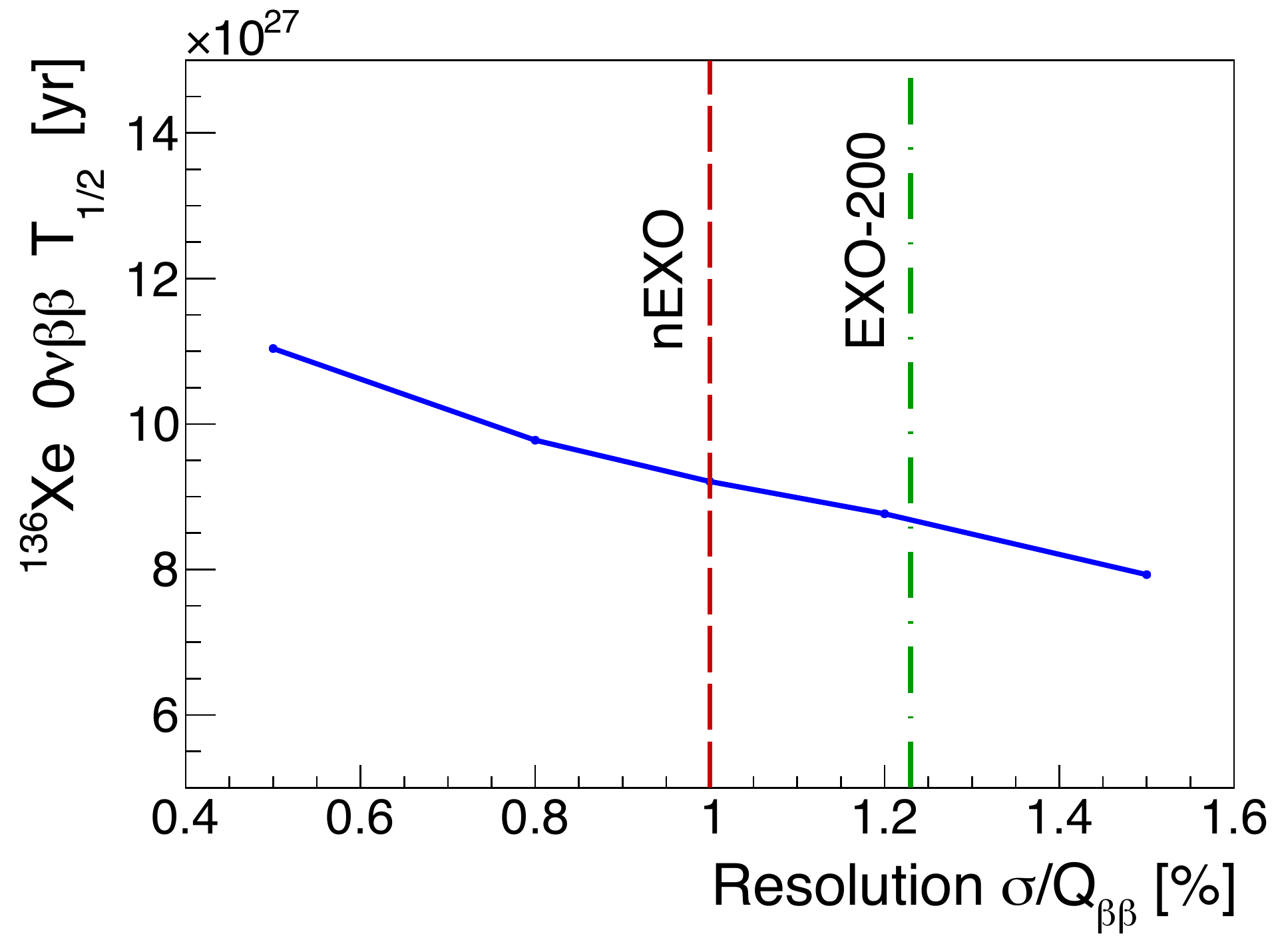}
\caption{nEXO median exclusion sensitivity at 90\% C.L. computed for different assumptions of the experiment's energy resolution.}
\label{fig:sen_vs_res}
\end{figure}

The impact of the energy resolution on nEXO's sensitivity is shown in Fig.~\ref{fig:sen_vs_res}. The shallow slope of this curve is understood by considering the role of the photoelectric peak  from \ce{^214Bi} background, which falls only 10 keV away from the \ce{^136Xe} \0 $Q$ value. In the range of energy resolutions considered here, only a small fraction of SS \ce{^{214}Bi} backgrounds lies more than $\pm$FWHM/2 away from \Q. Even at an energy resolution of 0.35\%, 50\% of the \ce{^{214}Bi} SS background falls within \Q$\pm$FWHM/2. 
For this reason, the half-life sensitivity does not significantly change with the energy resolution.  
On the other hand, the sub-dominant contribution arising from the fraction of SS \ce{^{208}Tl} decays that enter the same energy window is only $2.6\times10^{-5}$ at 1\% resolution. This fraction increases rapidly as the resolution worsens, becoming $2.8\times10^{-2}$ at $\sigma/\Q \sim1.5$\%.

\begin{figure}[tbp]
\centering
\includegraphics[width=0.95\columnwidth]{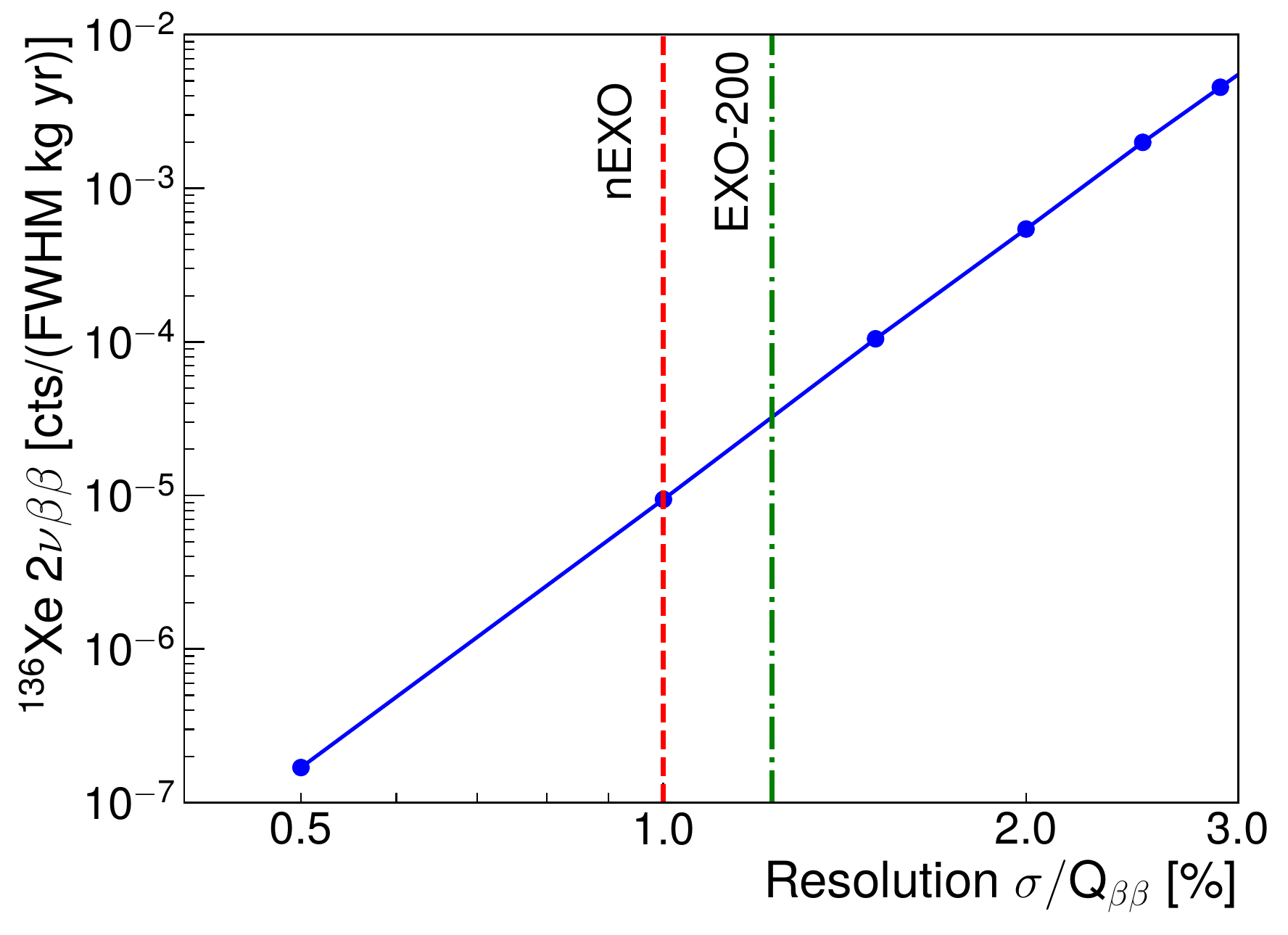}
\caption{Calculated \ce{^{136}Xe} \2 SS events falling within $\pm$ FWHM/2 of the \ce{^136Xe} Q-value as a function of the energy resolution (assumed Gaussian). The expected and measured energy resolution of nEXO and EXO-200 respectively are shown for reference.}
\label{fig:cts2n_vs_res}
\end{figure}

The distance to surface and event type parameters in nEXO's multiparameter analysis have no discriminating power against the unavoidable \2 background. As a result, energy resolution is the only proven method to suppress this background. Figure \ref{fig:cts2n_vs_res} shows the calculated \2 event rate in nEXO as a function of the energy resolution (assumed Gaussian). At nEXO's design energy resolution, the contribution of \2 decays at \Q$\pm$FWHM/2 amounts to only 0.34 counts over 10 years of data taking in the entire LXe volume, and is therefore negligible. This is also due to the fact that the \2 half-life for \ce{^136Xe} has been found to be larger than that of all other common \0 candidates \cite{Barabash:2015eza}. 

These results support nEXO's target energy resolution of $\sigma/\Q=1$\% and suggest that further improvements, while beneficial, are not critical to achieving a compelling sensitivity.
  
\begin{figure}[tbp]
\centering
\includegraphics[width=0.95\columnwidth]{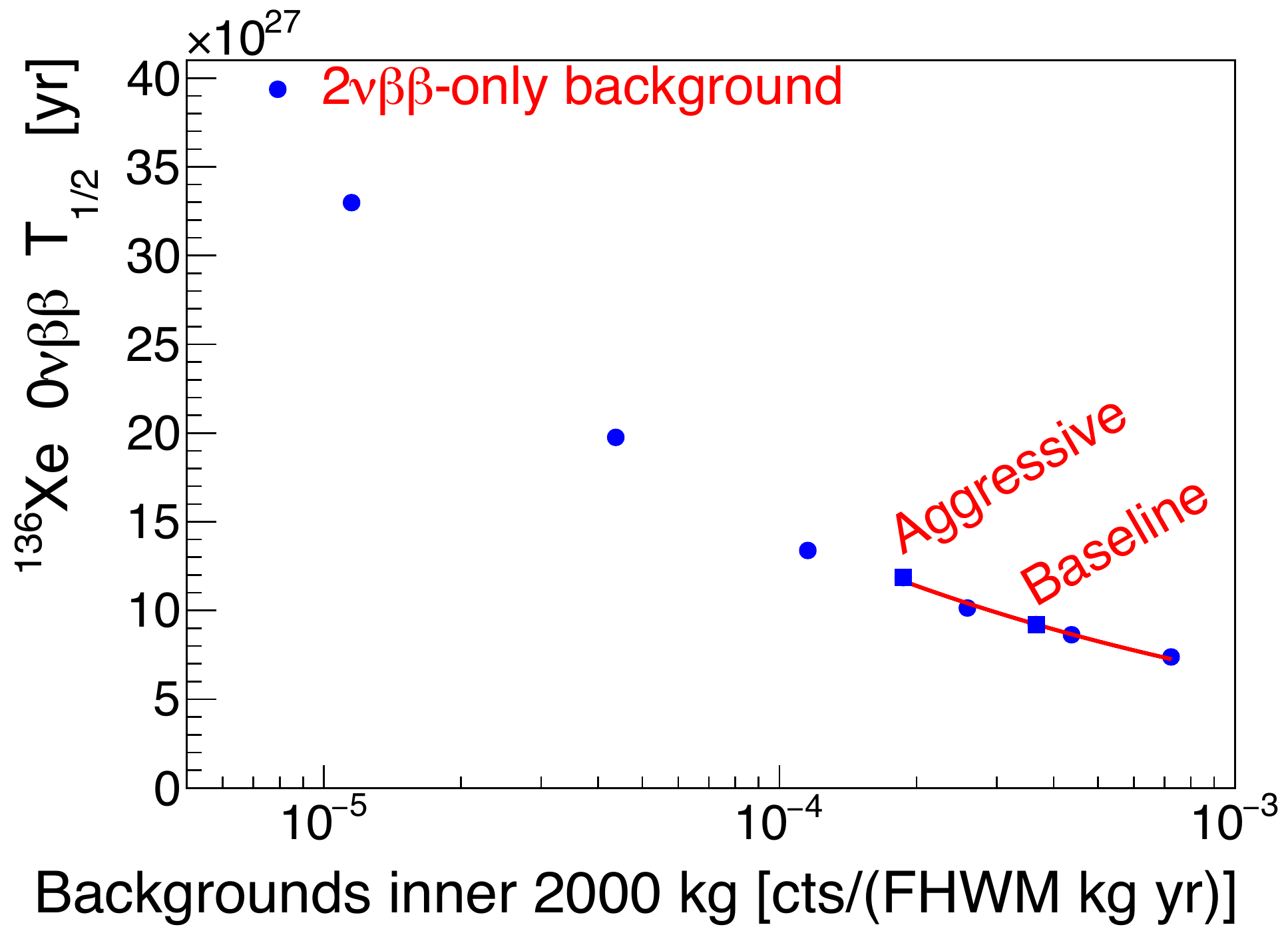}
\caption{Sensitivity (blue circles) to the \0 half-life of a nEXO-like experiment as a function of total background in \Q$\pm$FWHM/2 in the inner 2000 kg.  All components of nEXO's background model except for the \2 term are scaled to generate this curve. The red curve is the result of fitting the computed values with $T^{0\nu}_{1/2}\propto B^x$, giving $x=-0.35$.
}
\label{fig:sen_vs_bi}
\end{figure}

Ongoing efforts focus on reducing the SS backgrounds through advancement in material screening and selection, optimization of the detector components (e.g. mass and location), and improved analysis. A parametric study was performed to evaluate the improvement in \0 sensitivity as a function of the total background. All materials activities from table \ref{tab:rad} were uniformly scaled down by progressively larger fraction, with the exception of the \2 component which was held constant. New toy data sets were  generated and then fit to obtain a median sensitivity estimate for different background scenarios. 
The resulting curve is shown in Fig.~\ref{fig:sen_vs_bi}, assuming 10 years of data taking. The \0 sensitivity increases by a factor 4 as the background rate is lowered by two orders of magnitude. 
The point labeled ``baseline'' refers to the case described in this paper, while ``aggressive'' refers to a case in which still plausible improvements are made.

It is interesting to observe that for nEXO, the common approximation that sensitivity $T^{0\nu}_{1/2}$ scales with background $B$ as $1/\sqrt{B}$ is not valid. Indeed, fitting the calculated sensitivity points in Fig.~\ref{fig:sen_vs_bi}  with a power law results in 
$$
T^{0\nu}_{1/2}\propto \frac{1}{B^{0.35}}  
$$
This finding is significant. First, it underlines the importance of using experiment-specific techniques to estimate sensitivities. Second, it shows that nEXO is less sensitive to background fluctuations than what might be inferred from a simple $1/\sqrt{B}$ scaling. 
 
Hypothetically, the detection of a discrete energy deposit at $Q_{\beta\beta}$, due to some yet unknown decay, could lead to an unjustified claim of discovery of $0\nu\beta\beta$-decay. Clearly, having multiple competitive $\beta\beta$-experiments world-wide, based on different decaying nuclides, would provide a robust defense against this problem.
However, due to its multiparameter measurement capability nEXO is robust against such an unknown background. 
Discrete energy deposits are due to $\alpha$-particles, conversion electrons or $\gamma$-rays and can arise from decays internal or external to the LXe. 
An $\alpha$-decay could be identified by its characteristic scintillation to ionization ratio and is therefore not problematic for nEXO. The case of an unknown external $\gamma$-ray source can be  studied with the existing simulations. Such background would follow the same distribution in distance to surface and SS/MS as other external backgrounds. If an unknown decay were strong enough to produce as many SS events in the inner 3000 kg as a $3\sigma$ discovery at a half-life of $5.7\times10^{27}$ yr,
this decay would produce 271 counts in the MS outer volume, enough to rule out the expected background model at $p\; <\; 0.00001$. 
Due to their complexity, the other cases of unknown backgrounds from conversion electron or  $\gamma$-rays within the LXe will be the subject of future studies.

\begin{figure}[tbp]
\centering
\includegraphics[width=0.95\columnwidth]{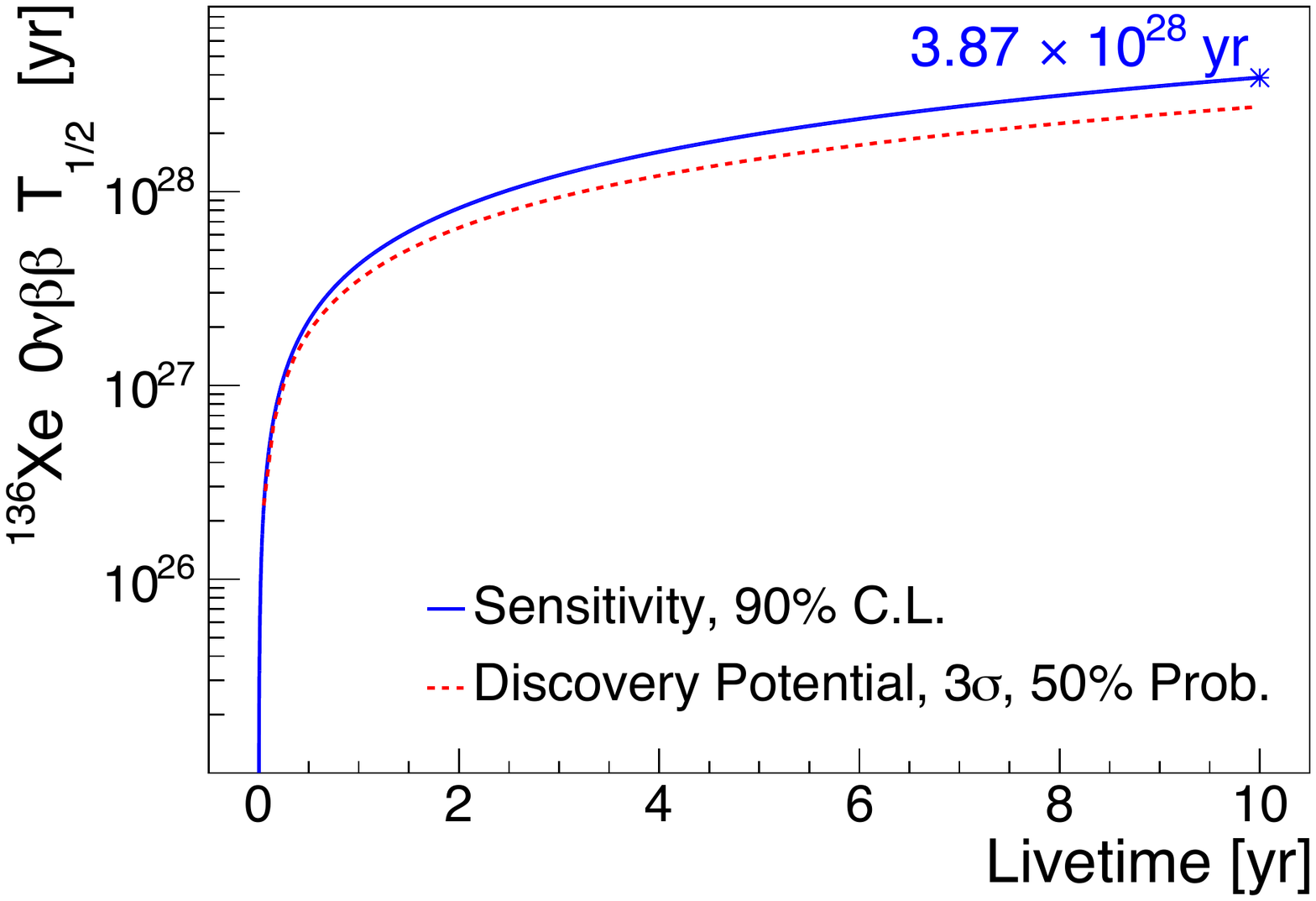}
\caption{Median exclusion sensitivity at 90\% C.L. and $3\sigma$ discovery potential to the \0 half-life of a nEXO-like experiment under a background consisting only of the \2 component. }
\label{fig:senBaTag}
\end{figure}

The case of an experiment entirely dominated by \2 background was considered as one limit to the possible sensitivity of nEXO. 
Techniques have been proposed that could potentially make this possibility a reality \cite{Twelker:2014zsa,Mong:2014iya}. This is an area of intense R\&D and, while challenges are still to be overcome, nEXO's current design leaves open the possibility of deploying upgrades at later stages of the experiment's lifetime. Sensitivity and discovery potential to \0 under a \2-only background scenario are provided in Fig.~\ref{fig:senBaTag}, in analogy with Fig.~\ref{fig:nom_sen}.

\section{Conclusions}

nEXO's sensitivity reach is rooted in the success and  experience gained from its predecessor EXO-200 and in the advantages provided by a large homogeneous TPC with good energy resolution, position reconstruction, and ability to identify particle type.

Making conservative assumptions on the detector and analysis performance, and using only measured radioassay inputs to build the background model, we predict that nEXO will reach a 3$\sigma$ discovery potential of $5.7\times10^{27}$ yr  for the \ce{^136Xe} \0 half-life. We further estimate its  90\% C.L. exclusion sensitivity to reach $9.2\times10^{27}$ yr.  Under  aggressive but not unrealistic assumptions, nEXO might reach well beyond a sensitivity of $10^{28}$ years.

\begin{figure}[tbp]
\centering
\includegraphics[width=\columnwidth]{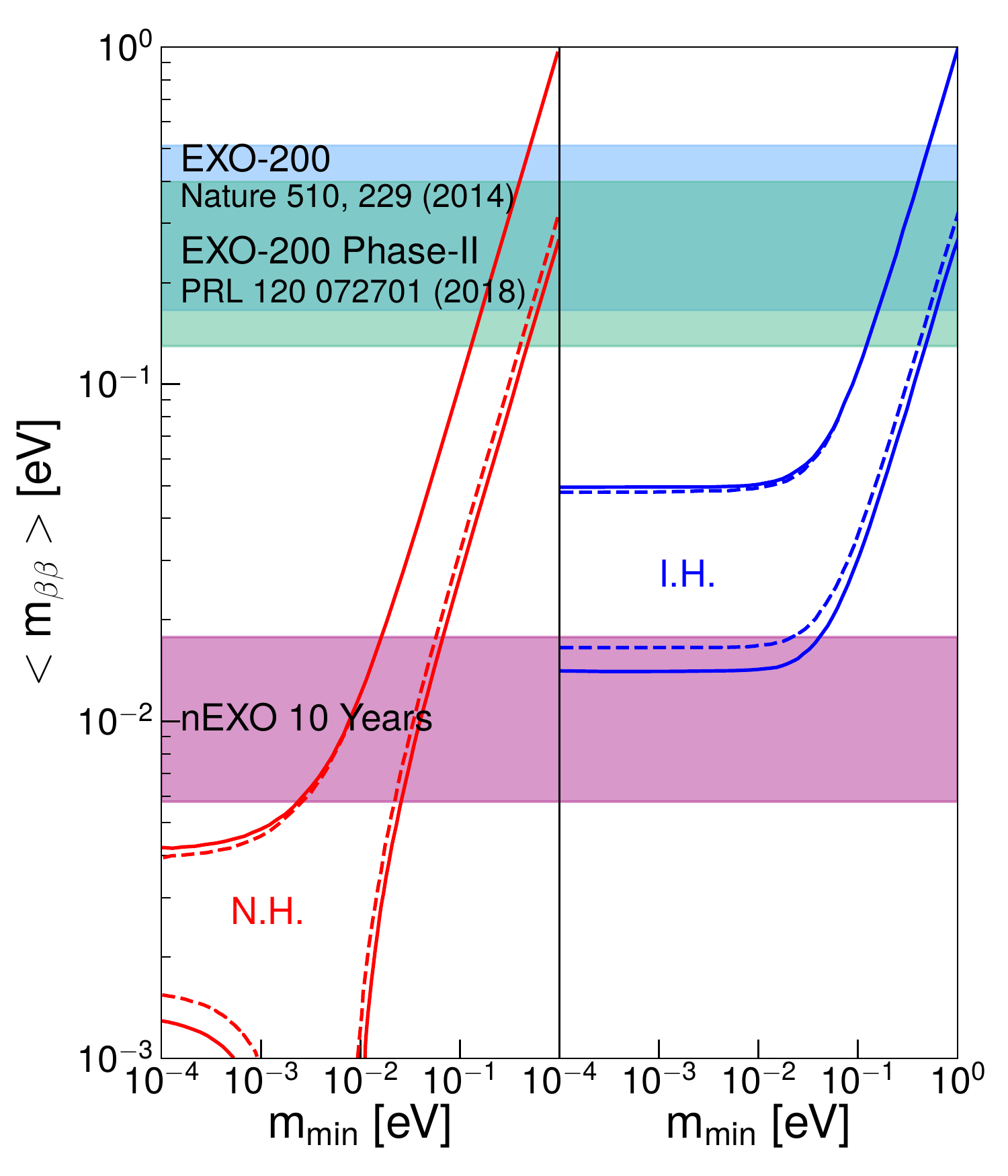}
\caption{90\% C.L. exclusion sensitivity reach to the effective majorana neutrino mass $m_{\beta\beta}$ as a function of the lightest neutrino mass for normal (left) and inverted (right) neutrino mass hierarchy. The width of the horizontal bands derive from the uncertainty in nuclear matrix elements (see text) and it assumes that $g_A = 1.27$. The width of the inner dashed bands result from the unknown Majorana phases and is irreducible. The outer solid lines incorporate the 90\% C.L. errors of the three-flavor neutrino fit of Ref.~\cite{Forero-PhysRevD.90.093006,Tortola:comm}.}
\label{fig:mbb_sensitivity}
\end{figure}

The sensitivity to the \0 half-life of \ce{^136Xe} can be converted into a reach on the effective Majorana neutrino mass $\langle m_{\beta\beta} \rangle$ under the assumption of a light Majorana neutrino exchange [Eq.~\eqref{eq:T-m}].  Figure \ref{fig:mbb_sensitivity} shows the nEXO exclusion sensitivity to $\langle m_{\beta\beta} \rangle$ as a function of the lightest neutrino mass. The allowed neutrino mass bands are derived from neutrino oscillation parameters from Refs.~\cite{Forero-PhysRevD.90.093006,Tortola:comm}. The $\langle m_{\beta\beta} \rangle$ exclusion band between 5.7 and 17.7 meV arises from the range of nuclear matrix elements, with  EDF \cite{PhysRevLett.111.142501} and QRPA \cite{PhysRevC.87.064302} at the minimum and maximum extreme respectively. A detailed evaluation of the sensitivity for various NME choices is given in Table~\ref{tab:mbb_nme_values}. Majorana neutrino masses are computed assuming an axial-vector coupling constant of $g_A = 1.27$ \cite{PDG2016}. 

Liquid xenon TPC is a proven, competitive technology in the search for \0.  3D event reconstruction in a monolithic detector at the tonne scale is a new powerful tool to reject, and perhaps more importantly identify, backgrounds; this is especially important for a discovery-class experiment.  Xenon is available at the tonne scale and is easily enriched thus simplifying the design and reducing the cost.  The nEXO experiment is expected to increase the sensitivity to \0 by about two orders of magnitude over current experiments, and has the multiparameter dataset to make a convincing case for the discovery of the Majorana nature of the neutrino and the violation of lepton number.

\begin{table}[h!]
\centering
\begin{tabular}{lccc}
\hline
Calculation & Reference & NME  & $\langle m_{\beta\beta}\rangle$ \\
            &           & $M_{0\nu}$ & [meV]                        \\
\hline
IBM-2         & \cite{PhysRevC.91.034304}   & 3.05  & 9.0                            \\
Skyrme-QRPA         & \cite{PhysRevC.87.064302}   & 1.55  & 17.7                            \\
QRPA        & \cite{PhysRevC.87.045501}   & 2.46   & 11.1					\\
RQRPA-UCOM        & \cite{JPG.39.12.2012.124006}   & 2.54  & 10.8                             \\
NREDF         & \cite{PhysRevLett.111.142501}   & 4.77 & 5.7                             \\
REDF        & \cite{PhysRevC.91.024316}   & 4.32  &  6.3                            \\
ISM         & \cite{NucPhysA.818.139.2009}   &  1.77  & 15.5                             \\
\hline
\end{tabular}
\caption{Nuclear Matrix Elements (NME) values and corresponding 90\% C.L. exclusion sensitivity limits on the Majorana neutrino mass $\langle m_{\beta\beta} \rangle$ for nEXO after 10 years of data taking. The values are computed for the \ce{^136Xe} \0 half-life sensitivity of $9.2\times10^{27}$ yr. (R)QRPA: (Renormalized) Quasi Random Phase Approximation; ISM: Interacting Shell Model; IBM: Interacting Boson Mode; (N)REDF: (Non)Relativistic Energy Density Functional. Majorana neutrino masses are computed assuming $g_A = 1.27$ \cite{PDG2016}.}
\label{tab:mbb_nme_values}
\end{table} 
 
\FloatBarrier

\section*{Acknowledgments}

We gratefully acknowledge the support of SNOLAB for the gamma and radon assay work conducted on site. We thank Mitchell Negus for contributing to the simulation work. 
This work has been supported by the Offices of Nuclear and High Energy Physics within DOE's Office of Science, and NSF in the United States; by NSERC, CFI, FRQNT, NRC, and the McDonald Institute (CFREF) in Canada; by SNF in Switzerland; by IBS in Korea; by RFBR in Russia; and by CAS and ISTCP in China. 
This work was supported in part by Laboratory Directed Research and Development (LDRD) programs at Brookhaven National Laboratory (BNL), Lawrence Livermore National Laboratory (LLNL), Oak Ridge National Laboratory (ORNL), and Pacific Northwest National Laboratory (PNNL).

\bibliography{nexo_sens.bib}

\end{document}